%% file: HusamWSN_6.tex
\documentclass{easychair}
\usepackage{supertabular}
\usepackage{graphicx}
\usepackage{longtable}
\usepackage{amsmath}
\usepackage{tabularx}

\usepackage{caption}
\usepackage{subcaption}

\DeclareUnicodeCharacter{2212}{-}   % added it: be careful of it

\usepackage{chronology}
\usepackage{algorithmic}
\usepackage[]{algorithm2e}

\usepackage{amssymb}
\usepackage{float}

\usepackage{threeparttable}

\usepackage{array}
\usepackage{subcaption}
\usepackage{multirow}
\usepackage{rotating}
\usepackage{pdflscape}

\usepackage{amssymb}
\begin{document}

%% Front Matter
%%
% Regular title as in the article class.
%
\title{$P\!\!\!_c\varepsilon\kappa_{\text{\small{max}}}$-Means++: Adapt-$P$ Driven by Energy and Distance Quality Probabilities Based on $\kappa$-Means++ for the Stable Election Protocol (SEP)}   %%%   CITATION OF YOUR PAPERS

% \titlerunning{} has to be set to either the main title or its shorter
% version for the running heads. Use {\sf} for highlighting your system
% name, application, or a tool.
%
\titlerunning{$P\!\!\!_c\varepsilon\kappa_{\text{\small{max}}}$-Means++: Adapt-$P$ Driven by $P\!\!\!_{\eta(j,i)}$ and $P\!\!\!_{\psi(j,i)}$ Quality Probabilities}

\volumeinfo
	{\emph{Journal of Wireless Mobile Networks, Ubiquitous Computing, and Dependable Applications}}
	{11}                        % volume
	{1 (Month)}              % issue
	{11}                         % starting page number

%% For only the editors. Authors, please keep this commented out
%\volumeinfo
%	{G. Sutcliffe, A. Voronkov} % editors
%	{2}                         % number of editors
%	{{\easychair} 1.0, 2008}    % event
%	{1}                         % volume
%	{1}                         % issue
%	{1}                         % starting page number

% Authors are joined by \and and their affiliations are on the
% subsequent lines separated by \\ just like the article class
% allows.

\author{\\
Husam Suleiman${}^{}$\thanks{Corresponding author: Department of Computer Engineering, Faculty of Computer and Information and Technology, Jordan University of Science and Technology, P.O. Box 3030, Irbid 22110, Jordan
}\\
${}^{}$Jordan University of Science and Technology, Irbid, Jordan\\
\url{hasuleiman@just.edu.jo}, Orcid: https://orcid.org/0000-0003-3805-7603
}

%${}^{}$Department of Computer Engineering, Faculty of Computer and Information Technology\\ Jordan University of Science and Technology, P.O. Box 3030, Irbid 22110, Jordan\\
%\url{hasuleiman@just.edu.jo}, Orcid: https://orcid.org/0000-0003-3805-7603
%}

%original
%\author{\\
%Tomoyuki Ishida${}^{1}$\thanks{Corresponding author: Department of of Computer and Information Sciences, Ibaraki University, 4121 Nakanarusawacho, Hitachi, Ibaraki, 3168511, Japan, Tel: +81-294-38-5138
%}~, Kazuhiro Takahagi${}^{1}$, Akira Sakuraba${}^{2}$, Noriki Uchida${}^{3}$, and Yoshitaka Shibata${}^{2}$\\
%${}^{1}$Ibaraki University, Hitachi, Ibaraki 3168511 Japan\\
%ishida@mx.ibaraki.ac.jp, 14nm712y@vc.ibaraki.ac.jp\\
%${}^{2}$Iwate Prefectural University, Takizawa, Iwate 0200693 Japan\\
%g236k001@s.iwate-pu.ac.jp, shibata@iwate-pu.ac.jp\\
%${}^{3}$Saitama Institute of Technology, Fukaya, Saitama 3692093 Japan\\
%uchida@sit.ac.jp
%}

%
%\author{Serguei A. Mokhov\thanks{Did all the difficult work}\\
%Concordia University\\
%Montreal, Quebec, Canada\\
%\url{mokhov@encs.concordia.ca}\\
%\and
%Geoff Sutcliffe\thanks{Did numerous tests and provided a lot of suggestions}\\
%University of Miami\\
%Miami, Florida, U.S.A.\\
%\url{geoff@cs.miami.edu}\\
%\and
%Andrei Voronkov\thanks{Masterminded EasyChair}\\
%University of Manchester\\
%Manchester, U.K.\\
%\url{andrei@voronkov.com}\\
%}

% \authorrunning{} has to be set for the shorter version of the authors' names;
% otherwise a warning will be rendered in the running heads.
%

%\authorrunning{Mokhov, Sutcliffe, and Voronkov} % I can use it instead of below when I have multiple authors
\authorrunning{Husam Suleiman et al.}
%\authorrunning{}

\maketitle

%------------------------------------------------------------------------------
% Abstract
%
\begin{abstract}
\noindent

Attaining a prolonged network lifetime, maximized coverage, and high performance are vital design factors that have to be maintained in a Wireless Sensor Networks (WSN). Such factors are dependent on the stability and optimality of the protocol employed to formulate Sensor Nodes (SNs) into mutual clusters that effectively work around fulfilling specific performance goals. SEP is a heterogeneity-aware protocol implemented based on Low-Energy Adaptive Clustering Hierarchy (LEACH) protocol, designed to prolong the stability period of the network defined by the time interval before the death of the first SN.
Nevertheless, adaptability in design is a scheme that has not been extensively applied in the formation of WSN election protocols. Adapt-$P$ is a probabilistic model for adaptivity designed to evolve the probability of selecting a cluster-head based on the active status of the WSN, represented by the residual energy of and distances between SNs. However, the adaptive probability $P\!\!\!_{\text{\tiny{adp}}}$ formalized in Adapt-$P$ is developed based on the remaining number of SNs $\zeta$ and optimal clustering $\kappa_{\text{\tiny{max}}}$, yet $P\!\!\!_{\text{\tiny{adp}}}$ does not implement the probabilistic ratios of energy and distance factors in the network. Furthermore, Adapt-$P$ does not localize cluster-heads in the first round properly because of its reliance on distance computations defined in LEACH, that might result in uneven distribution of cluster-heads in the WSN area and hence might at some rounds yield inefficient consumption of energy.
This paper utilizes \nolinebreak{$k$\small{-}means\small{++}} and Adapt-$P$ to propose \nolinebreak{$P\!\!\!_{\text{c}} \kappa_{\text{\tiny{max}}}$\small{-}means\small{++}} clustering algorithm that better manages the distribution of cluster-heads and produces an enhanced performance. The algorithm employs an optimized cluster-head election probability $P\!\!\!_\text{c}$ developed based on energy-based $P\!\!\!_{\eta(j,i)}$ and distance-based $P\!\!\!_{\psi(j,i)}$ quality probabilities along with the adaptive probability $P\!\!\!_{\text{\tiny{adp}}}$, utilizing the energy $\varepsilon$ and distance optimality $d\!_{\text{\tiny{opt}}}$ factors. Furthermore, the algorithm utilizes the optimal clustering $\kappa_{\text{\tiny{max}}}$ derived in Adapt-$P$ to perform adaptive clustering through \nolinebreak{$\kappa_{\text{\tiny{max}}}$\small{-}means\small{++}}. The proposed \nolinebreak{$P\!\!\!_{\text{c}} \kappa_{\text{\tiny{max}}}${\small{-}}means{\small{++}}} is compared with the energy-based algorithm \nolinebreak{$P\!\!\!_\eta \varepsilon \kappa_{\text{\tiny{max}}}${\small{-}}means{\small{++}}} and distance-based \nolinebreak{$P\!\!\!_\psi d_{\text{\tiny{opt}}} \kappa_{\text{\tiny{max}}}${\small{-}}means{\small{++}}} algorithm, and has shown an optimized performance in term of residual energy and stability period of the network.

\end{abstract}

%------------------------------------------------------------------------------

\input{WSN_Introduction}

\input{WSN_Contributions}

\input{WSN_Background}

\input{WSN_NetworkModel}

\input{WSN_MathModel}

\input{WSN_Evaluation}

\input{WSN_Conc_Future}

%------------------------------------------------------------------------------

\section*{Acknowledgement}
This research has been approved and documented in the Deanship of Research at Jordan University of Science and Technology under the Research Grant Number: 20230497.

% was supported

% Refs:
%
\label{sect:bib}
\bibliographystyle{abbrv}
\bibliography{easychairWSN}

\section*{Author Biography}
\vspace*{1em}
%------------------------------------------------------------------------------
\begin{biography}{Husam Suleiman}{husam} received his PhD in electrical and computer engineering from University of Waterloo, Canada in 2019. His MSc degree is in computer engineering from Khalifa University, UAE in 2011 in collaboration with the Massachusetts Institute of Technology (MIT). His BSc degree is in electrical and computer engineering from Hashemite University, Jordan in 2007. Currently, Dr. Suleiman is an assistant professor in Jordan University of Science and Technology, Irbid, Jordan. His research interests include QoS optimization in multi-tier cloud computing, load scheduling \& balancing, big data algorithms, resource allocation methods, performance prediction analysis, and security requirements engineering methods for smart grids and systems.
\end{biography}
%%------------------------------------------------------------------------------
%\vspace*{0.5em}
%%------------------------------------------------------------------------------

%------------------------------------------------------------------------------
%\appendix
%\section{{\easychair} Requirements Specification}
%\label{sect:easychair-requirements}
%
%The following high-level requirements were set for the development of
%the {\easychair} class, and were refined further development went along.

%------------------------------------------------------------------------------
\end{document}

%% file: WSN_Introduction.tex
\section{Introduction}

%\PARstart{T}{he}
The extensive usage of WSNs in various real-time applications has recently paid researcher attentions toward developing effective optimization mechanisms to achieve an improved performance and design goals of the network~\cite{gouda2020systematic, Zagrouba21, ARJUNAN2019, agrawal22}. A WSN is generally defined as a large-scale network environment with many low-powered SNs that require efficient schemes for data aggregation and transmission~\cite{Ogundile, Jie, Mung, kaur21}. Controlling energy consumption, scalability, reliability, and topology-based difficulties are primary challenges that should be considered when such mechanisms are developed~\cite{alomari2022systematic, Jukuntla22}.

A WSN suffers from limited computational, communication, and energy capabilities that lower the lifetime and coverage of SNs~\cite{wu2022dual, smail2017energy, Pushpa}. Such nodes are often deployed in areas where energy is a constraint~\cite{razaque2016h, Juwaid2018, Kumawat}. It is typical that a WSN is divided into clusters to reduce the complexity of data transmission and achieve energy saving~\cite{bidaki2016towards, Zafor22, pitchaimanickam2020hybrid}, in which cluster-heads are elected to gather and send SN data to a Base Station (BS) for further processing and analysis. Each SN has a potential of becoming a cluster-head in each round, where energy is among the multiple factors that play the role of such election process. Hence, SN deployments and clustering mechanisms are primary concerns to achieve a maximum performance.

Placement scenarios of SNs often involve random deployments in the sensor field. Such placements make it difficult to formulate clusters that are required to effectively cover the field and account for performance goals of the WSN. Models in existing literature elect cluster-heads randomly in the first round. The election process in subsequent rounds chooses cluster-heads based on a criterion that typically considers conflicting factors such as residual energy of SNs, number of cluster-heads permitted per round, and probability of a SN to become a cluster-head. Therefore, selection of cluster-heads is an optimization problem that plays a critical role in improving energy efficiency of SNs and network lifetime.

\section{Motivation}

Adaptability in design is a paradigm that has recently been involved in formulating communication models of WSNs. It is essential that a network is designed such that it is adaptive to state variations and operational contexts. Adapt-$P$~\cite{suleiman2022cost} is a probabilistic model that employs adaptivity in the probability of selecting a cluster-head. An adaptive function is designed to dynamically evolve the probability based on the changed status of the WSN represented by the maximum number of cluster-heads permitted per round, residual energy of SNs, and number of alive SNs remained in the WSN.

Nonetheless, effectively initializing placements of cluster-heads in the first round facilitates the convergence toward distributing the role of cluster-head fairly among SNs in subsequent rounds, and most importantly helps effectively cover the WSN field throughout the network running-time. Such localizations are required to avoid depleting energy of a SN while other SNs are not. The \nolinebreak{$\kappa$\small{-}means} is a clustering algorithm that divides the space of nodes into $K$ clusters using the squared Euclidian distances. The main drawback of \nolinebreak{$\kappa$\small{-}means} is that localizing a centroid in a far place may end-up having no SNs associated with it. The \nolinebreak{$\kappa$\small{-}means\small{++}} built upon the standard \nolinebreak{$\kappa$\small{-}means} algorithm is developed to overcome such drawbacks by controlling the initialization of clusters~\cite{kmeans}. The \nolinebreak{$\kappa$\small{-}means\small{++}} selects the first cluster-head uniformly at random in the first round, after which cluster-heads are selected with a probability based on the distance between each SN and the chosen cluster-head.

Making active run-time clustering decisions in WSNs is therefore essential to maintain a stable performance. It is required that SNs are fairly elected to act as cluster-heads such that the WSN field is uniformly covered and the energy is preserved to maximize the network's lifetime. The adaptive probability $P\!_{\text{\tiny{adp}}}$ computed in Adapt-$P$ manages the election process according to the evolved network status and thus highlights the need to investigate its usage within the \nolinebreak{$\kappa$\small{-}means\small{++}} algorithm to derive a probabilistic model and formulate clusters that efficiently account for such conflicting objectives.

%% file: WSN_Contributions.tex
\section{Contributions}
\label{sec:contr}

An adaptable design is vital to cope with changes and variations occurred in the context of the networking environment. This paper presents the \nolinebreak{$P\!\!\!_{\text{c}} \kappa_{\text{\tiny{max}}}$\small{-}means\small{++}} clustering algorithm that produces an improved performance based on \nolinebreak{$k${\small{-}}means{\small{++}}} and Adapt-$P$ algorithms. Contributions of this paper are summarized as follows:
\begin{itemize}
  \item The optimal clustering $\kappa_{\text{\tiny{max}}}$ derived in Adapt-$P$ is employed in the context of \nolinebreak{$k${\small{-}}means{\small{++}}} to adaptively control the formulation of the maximum number of clusters permitted per round, producing \nolinebreak{$\kappa_{\text{\tiny{max}}}${\small{-}}means{\small{++}}}.

  \item Energy-based \nolinebreak{$P\!\!\!_\eta \varepsilon \kappa_{\text{\tiny{max}}}${\small{-}}means{\small{++}}} and distance-based \nolinebreak{$P\!\!\!_\psi d_{\text{\tiny{opt}}} \kappa_{\text{\tiny{max}}}${\small{-}}means{\small{++}}} clustering algorithms are respectively developed based on energy-based $P\!\!\!_{\eta(j,i)}$ and distance-based $P\!\!\!_{\psi(j,i)}$ quality probabilities, derived by utilizing the energy $\varepsilon$ and distance optimality $d\!_{\text{\tiny{opt}}}$ factors.

  \item A clustering algorithm \nolinebreak{$P\!\!\!_{\text{c}} \kappa_{\text{\tiny{max}}}${\small{-}}means{\small{++}}} is developed based on a cluster-head election probability $P\!\!\!_{\text{c}}$ derived by utilizing the adaptive probability $P\!\!\!_{\text{\tiny{adp}}}$ formalized in Adapt-$P$ along with the energy-based $P\!\!\!_{\eta(j,i)}$ and the distance-based $P\!\!\!_{\psi(j,i)}$ quality probabilities.
\end{itemize}

In that, the deployment and clustering strategy is driven by adaptive decisions triggered by the election probability $P\!\!\!_{\text{c}}$ to reform clusters based on \nolinebreak{$\kappa_{\text{\tiny{max}}}${\small{-}}means{\small{++}}}.

%% file: WSN_Background.tex
\section{Background and Related Work}
\label{sec:background}

Determining effective clustering and scheduling procedures for SNs are primary concerns in WSN so that network's lifetime and QoS requirements are optimized~\cite{mahboub2017energy, wu2022dual, mechta2014leach}. Such procedures simplify the complexity and management of the network especially when its size increases~\cite{Pundir20, Ismail20}. Existing works in the literature are mainly focused on LEACH-based and SEP-based algorithms that aim at prolonging the lifetime by producing energy-efficient clusters~\cite{Reddy2023, SinghQoS2022, Azzouz2022, Daogen2022}. The LEACH protocol proposed by Heinzelman et al.~\cite{Heinzelman, Heinzelman2} randomly rotates the role of cluster-head among SNs to maximize their lifetimes and area of coverage. The mathematical model presented in LEACH computes the number of cluster-heads required to generate clusters, in which the LEACH algorithm outperforms static and Minimum Transmission Energy (MTE) algorithms. The SEP protocol proposed by Smaragdakis et al.~\cite{SEP_SANPA04} accounts for heterogeneity of SNs by deploying nodes with different initial residual energy, in which they aim at prolonging the stability period defined by the time to death of the first SN.

It is critical to assess the state of a network before a decision is made~\cite{suleiman2022cost}. Adaptability is employed in our previous work Adapt-$P$~\cite{suleiman2021adaptP} to evaluate the network state, in which the probability of selecting a cluster-head is dynamically adapted according to state variations at run-time. Such a state is represented by the network structure, conditions, and percentage of alive SNs. A near-optimal distance between each cluster-head and its associated cluster-members is formulated based on LEACH such that the total energy consumption is mitigated.

Existing work in the literature presents different models and mechanisms for energy management and lifetime enhancements in WSNs~\cite{Jamshed22, Junqi2022, Humaira2022, Siwen2023}. For instance, Gawade et al.~\cite{gawade2016centralized} present a centralized energy-efficient distance-based routing to distribute energy uniformly between SNs. A distributed algorithm is proposed to select cluster-heads based on SN's energy consumption and distance to the BS. Han et al.~\cite{han2022novel} propose an energy-efficient clustering protocol for the energy harvesting-based WSNs. They attempt to maximize the network coverage by a control mechanism that adjusts the number of SNs permitted to enter the energy-harvesting mode and the data transmission mode. Panchal et al.~\cite{panchal2020rch} balance the load between clusters by calculating an energy-based threshold that identifies a SN as a cluster-head if its residual energy factor is greater than the threshold. Lim et al.~\cite{lim2011adaptive} present a WSN scheduling algorithm that adaptively forms sampling schedules for data gathering by finding temporal and spatial data correlations.

The \nolinebreak{$\kappa$\small{-}means} algorithm has also been utilized to solve clustering problems in WSNs~\cite{mishra2019trust, periyasamy2016balanced, ben2022energy, kandari2014k}. For example, Periyasamy et al.~\cite{periyasamy2016balanced} present a modified \nolinebreak{$\kappa$\small{-}means} algorithm that divides the space into clusters, each of which entails three cluster-heads simultaneously. The algorithm uses a mechanism to share the network load between the three cluster-heads of a cluster by having only one cluster-head active at a time. In this scenario, the algorithm reduces the number of times the clustering procedure has to be performed, increases the percentage of data packets transmitted to the BS, and reduces the energy consumption which as a result increases the network's lifetime.

Lehsaini et al.~\cite{lehsaini2018improved} propose a routing scheme based on \nolinebreak{$\kappa$\small{-}means} clustering that facilitates transmission and distribution of loads fairly among cluster-heads. They present a re-affiliation phase to formulate clusters that are equal in size in terms of number of SNs, after which they compute the average number of members within clusters. Then, the SN whose cluster's size is above the average joins a cluster at its border whose size is below the average. Li et al.~\cite{lin2018efficient} propose an efficient routing protocol based on \nolinebreak{$\kappa$\small{-}means} and fuzzy logic, in which the sink node uses \nolinebreak{$\kappa$\small{-}means\small{++}} to divide the WSN into clusters and utilizes fuzzy rules to compute the value of SNs so that SNs with highest values are chosen to be cluster-heads. 

Wu et al.~\cite{wu2022dual} present a hybrid \nolinebreak{$\kappa$\small{-}means} and Canopy algorithms to produce energy-efficient routing protocol based on LEACH. The protocol aims at minimizing the load on the cluster-head by electing a vice cluster-head based on SN's residual energy and the distance from SNs to the BS. Gouissem et al.~\cite{ben2022energy} use \nolinebreak{$\kappa$\small{-}means} algorithm with grid-based routing for energy-efficient WSN clustering. An optimal size of a grid is calculated by the BS, after which the \nolinebreak{$\kappa$\small{-}means} is utilized to determine the cluster-head in each grid-cell. Kandari et al.~\cite{kandari2014k} present a $k$-SEP algorithm that utilizes the \nolinebreak{$\kappa$\small{-}means} to form clusters until $50\%$ of SNs die and then switch to random clustering.

Meta-heuristic mechanisms are also employed to solve problems in WSNs~\cite{gouda2020systematic, Shashi2018, pal2020eewc, zhang2017fuzzy, lata2020fuzzy}. Bhushan et al.~\cite{Shashi2018} combine the Genetic Algorithm (GA) and \nolinebreak{$\kappa$\small{-}means} clustering, in which the problem is formalized by finding an optimal number of clusters in a big search space of heterogeneous SNs such that energy is saved and network's lifetime is maximized. Also, Pal et al.~\cite{pal2020eewc} use the GA to produce clusters by modifying the set-up phase in LEACH. The algorithm defines a chromosome in a GA as a set of SNs available in the field. The fitness function of a chromosome depends on the cluster compactness, cluster separation, and normalized number of cluster-heads; with different weights to each factor. After evolving the GA, the chromosome with the minimum fitness value is selected to formalize clusters and their cluster-heads.

Lata et al.~\cite{lata2020fuzzy} discuss the problem of existing decentralized WSN clustering that might possibly form clusters with cluster-heads elected to be both at the borderline of their clusters and near to each other, which accordingly could result in poor energy saving and lifetime enhancement of SNs. To overcome such drawbacks, they adopt a fuzzy approach to cluster SNs using a centralized mechanism in which they elect a cluster-head and a vice-cluster-head. Zhang et al.~\cite{zhang2017fuzzy} employ a distributed scheme to cluster SNs based on a fuzzy approach to decide whether a SN is qualified to act as a cluster-head or not. The proposed algorithm considers residual energy of the SN assessed to be a cluster-head, as well as number and residual energy of neighbor SNs within a pre-defined transmission range.

Furthermore, the Particle Swarm Optimization (PSO) algorithm has been utilized in forming clusters in WSNs~\cite{guhan2021eedchs, rao2017particle, dohare2019pso, gamal2022enhancing}. For instance, Guhan et al.~\cite{guhan2021eedchs} apply the PSO to choose cluster-heads based on distance of transmission between SNs. Rao et al.~\cite{rao2017particle} develop a utility function employed in PSO to formulate clusters in which they account for residual energy of SNs, distances between SNs together, and distances of SNs to the sink node so that a cluster-member accurately joins a better cluster-head. In addition, Dohare et al.~\cite{dohare2019pso} propose an energy-efficient clustering protocol for Internet of Things (IoTs) that combines \nolinebreak{$\kappa$\small{-}means} and PSO algorithms to form clusters and select cluster-heads, respectively. Gamal et al.~\cite{gamal2022enhancing} propose a fuzzy logic LEACH-based technique to enhance the lifetime of a WSN. A hybrid PSO and \nolinebreak{$\kappa$\small{-}means} algorithms are utilized to create clusters, in which primary and secondary cluster-heads are selected using the fuzzy logic.

Overall, it is of paramount importance to evaluate states of the network and act adaptively at run-time so that energy is conserved and the lifetime is maximized with the least cost. In this paper, the set-up phase in a WSN is reformed so that clusters are actively evolved based on network states. The adaptive probabilistic model presented in Adapt-$P$~\cite{suleiman2021adaptP} is employed in \nolinebreak{$\kappa$\small{-}means\small{++}} to formulate clusters and accurately evolve the network state according to its existing conditions. The maximum number of clusters $\kappa_{\text{\tiny{max}}}$ optimized based on $d_{\text{\tiny{opt}}}$ is formalized in \nolinebreak{$\kappa$\small{-}means\small{++}}. The cluster-head selection probability $P\!\!_{\text{\tiny{adp}}}$ is adapted based on number of alive SNs $\zeta$, distances of cluster-members from their corresponding cluster-heads evolved in \nolinebreak{$\kappa$\small{-}means\small{++}}, and residual energy of SNs in the clusters. A probabilistic model is then derived to compute the probability value of a node to act as a cluster-head to accordingly compose clusters.

%% file: WSN_NetworkModel.tex
\section{Network Model}
\label{sec:NetMod}

The network model simulates a set $\mathbb{S}$ of $n$ stationary SNs distributed in a two-dimensional space of \nolinebreak{$a$x$a$ m$^2$} and supported with a single BS (sink) located in the middle of the field to process data, in which the BS is assumed to always have adequate computational and power supply for data transmission and analysis.
\begin{equation}
\label{equ:SNs}
   \mathbb{S} = \{1, 2, 3, ..., j, ..., n\},~~~~~~~~~\forall j\!\in[1,\!n]
\end{equation}
SNs in contrast are energy-constraint, yet the energy resided in each SN is sufficient to exchange data with other SNs and to reach the BS if the SN is required to act as a cluster-head.

SNs deployed in the network are assumed to be heterogeneous, each of which starts with a particular initial amount of energy. The model in SEP is adopted in this paper to create two types of SNs according to their initial energy, namely normal and advanced SNs. The percentage of SNs that are advanced is $\nu$. A normal SN $s\!_j$ starts with an initial energy-level $\nolinebreak{\varepsilon_{\text{\tiny{init}}}^j\!=\!\varepsilon_{0}^j}$ whilst an advanced SN starts with $\nolinebreak{\varepsilon_{\text{\tiny{init}}}^j\!=\!{\varepsilon_0^j}(1\!+\!b)}$, where $\nolinebreak{b\!\in\!\mathbb{Z}^{+}}$ is a positive constant integer and $\varepsilon_{0}^j$ is the initial amount of energy reserved in a SN $s\!_j$ at the time of deployment. Each cluster of SNs entails a single cluster-head and multiple cluster-members. Each cluster-member is associated with only one cluster-head. A cluster-member does not directly transfer data to the BS, instead the cluster-member communicates with the BS via its associated cluster-head during its Time Division Multiple Access (TDMA) period. The communication channel is symmetric in which the energy required to transmit data packets from a source SN to a destination SN and vise versa is the same, for a given signal-to-noise ratio.

\section{Energy Model}
\label{sec:EnergyMod}

The data packet exchanged between SNs together, as well as between any SN and the BS, is assumed to have a fixed size represented by $\ell$ bits. The energy model described by Heinzelman et al.~\cite{Heinzelman} derives the energy consumption for a SN per data packet of size $\ell$ bits, in which a SN consumes energy to run its radio electronics and power amplifier. It is formulated that the total energy consumption is modeled by the energy $\varepsilon_{\text{rx}}(\ell)$ required to receive, energy $\varepsilon_{\text{da}}(\ell)$ to aggregate, and energy $\varepsilon_{\text{tx}}(\ell,d)$ to transmit a data packet of size $\ell$ bits over a distance $d$ between source and destination SNs. Heinzelman et al.~\cite{Heinzelman} adopt two distance-based models: free-space and multi-path fading. The free-space model is employed if the distance $d$ between SNs exchanging data packets is less than a distance threshold $d_0$ (indicated by $\nolinebreak{d\!\leq\!d_0}$); otherwise the multi-path fading model is utilized ($\nolinebreak{d\!>\!d_0}$).

A transmitter SN consumes energy $\varepsilon_{\text{tx}}(\ell,d)$ when a data packet of size $\ell$ bits is transferred over a distance $d$ between source and destination SNs. Such energy includes the energy $\varepsilon_{\bar{e}}$ consumed per bit to run the transmitter's radio electronics, the energy factor $\epsilon_{\text{{fs}}}$ required to run the power amplifier in the free-space (fs) model, and the energy amplification factor $\epsilon_{\text{{mp}}}$ required to run the power amplifier in the multi-path fading (mp) model, as follows:
\begin{equation}
\label{equ:E_TX}
\varepsilon_{\text{tx}}(\ell,d) =
\begin{cases}
     \ell\varepsilon_{\bar{e}} + \ell\epsilon_{\text{fs}} d^2, \;\;\;\;\;\;\;\;\;\;  d \leq d_0  \\
     \ell\varepsilon_{\bar{e}} + \ell\epsilon_{\text{mp}} d^4, \;\;\;\;\;\;\;\;\;  d > d_0
\end{cases}
\end{equation}
where the threshold distance $d_0$ is represented by:
\begin{equation}
\label{equ:E_o}
    d_0 = \sqrt{\frac{\epsilon_{\text{fs}}}{\epsilon_{\text{mp}}}}
\end{equation}

For the threshold $d_0$, the distance $d$ between a source SN and a destination SN decides whether to employ the free-space (fs) model or the multi-path (mp) fading model. In this paper, the free-space model ($d^2$) is utilized for intra-cluster data transmission because the distance $d$ is assumed to be $d\!\leq\!d_0$, whereas the multi-path fading model ($d^4$) is used for data transmission between a cluster-head and the BS because the distance $d$ is assumed to be $d\!>\!d_0$.

A receiver SN consumes energy $\varepsilon_{\text{rx}}(\ell)$ to only run the radio electronics, that is independent of the distance $d$ between SNs and is dissipated to receive a data packet of size $\ell$ bits as follows:
\begin{equation}
\label{equ:E_RX}
    \varepsilon_{\text{rx}}(\ell) = \ell\varepsilon_{\bar{e}}
\end{equation}
where $\varepsilon_{\bar{e}}$ is the energy consumed per bit to run the receiver's radio electronics. The receiver as well consumes the energy $\varepsilon_{\text{da}}(\ell)$ to aggregate a data packet of size $\ell$ bits, that is equal to the receiving energy $\varepsilon_{\text{rx}}(\ell)$ computed in equation~\ref{equ:E_RX}.

\section{Notations}
\label{sec:nots}

Parameters of the network simulation are shown in Table~1. It illustrates the maximum number of rounds, field dimensions (in meters), SN parameters, values for SNs' heterogeneity, and parameters of the energy model.
\begin{table*}[!ht]
\label{tab:notations}
\scalebox{0.9}{
\begin{tabular}{c|l}
\hline
Notation                                & Definition           \\ \hline
$a$                                     & Dimension of the field \\
$b$                                     & Positive constant integer \\
$\mathbb{C}$                            & Set of clusters \\
$d_0$                                   & Distance threshold to distinguish between free-space and multi-path fading models \\
$d_{\text{bs}}$                         & Distance to BS \\
$d\!_{\text{\tiny{opt}}}$               & Optimal distance between a cluster-head and its cluster-members \\
$D(s\!_j, h_i)$                         & Euclidian distance between a SN $s\!_j$ and a cluster-head $h_i$ \\
$D(s\!_j, O_i)$                         & Euclidian distance between a SN $s\!_j$ and a centroid $C_i$ \\
$\varepsilon_{\text{\tiny{init}}}^j$    & Initial energy of $j^{\text{th}}$ SN \\
$\varepsilon_{\text{rx}}(\ell)$         & Energy required to receive a data packet of size $\ell$ bits \\
$\varepsilon_{\text{tx}}(\ell,d)$       & Energy required to transmit a data packet of size $\ell$ bits over a distance $d$ \\
$\varepsilon_{\text{da}}(\ell)$         & Energy required to aggregate a data packet of size $\ell$ bits over a distance $d$ \\
$\varepsilon_{\bar{e}}$                 & Energy consumed per bit to run the transmitter and receiver radio electronics \\
$\varepsilon_{\text{\tiny{res}}}^{(j,i)}$ & Residual energy-level of SN $s\!_j$ in a cluster $C_i$ \\
$\varepsilon_{\text{\tiny{init}}}^{(j,i)}$ & Initial energy-level of SN $s\!_j$ in a cluster $C_i$ \\
$\varepsilon_{\text{\tiny{avg}}}^i$         & Average energy-level of a cluster $C_i$ \\
$\epsilon_{\text{{fs}}}$        & Energy factor required to run the power amplifier in the free-space (fs) model \\
$\epsilon_{\text{{mp}}}$        & Energy amplification factor required to run the power amplifier in the multi-path fading (mp) model \\
$h_i$                           & $i^{\text{th}}$ cluster-head \\
$\mathbb{H}$                    & Set of cluster-heads \\
$i$                             & ID of a cluster \\
$j$                             & ID of a SN \\
$\kappa_{\text{\tiny{max}}}$    & Maximum number of clusters per round \\
$\ell$                          & Size of data packets in bits \\
$n$                             & Total number of SNs in the field \\
$N_i$                           & Number of SNs within cluster $i$ \\
$\eta(j,i)$                     & Energy coefficient of SN $s\!_j$ in a cluster $C_i$ \\
$O_i$                           & Centroid of a cluster $C_i$ \\
$P\!\!\!_{\sigma(j,i)}$         & Deviation-based distance quality probability of a SN $s\!_j$ in a cluster $C_i$ \\
$P\!\!\!_{\text{\tiny{adp}}}$   & Adaptive probability of selecting a cluster-head \\
$P\!\!\!_{\text{c}}$            & Cluster-head selection probability \\
$P\!\!\!_{\eta(j,i)}$           & Energy-based election probability \\
$P\!\!\!_{\psi(j,i)}$           & Distance-based election probability \\
$P\!_{\xi(j,i)}$                & Global quality probability for SN $s\!_j$ in cluster $C_i$ \\
$P\!\!\!_{\varrho(j,i)}$        & Local quality probability for SN $s\!_j$ in cluster $C_i$ \\
$P\!\!_{\gamma(j,i)}$           & Optimal-oriented distance quality probability of a SN $s\!_j$ \\
$P\!\!\!_{\sigma(j,i)}$         & Deviation-based distance quality probability of SN $s\!_j$ in cluster $C_i$ \\
$\Delta_{i}$                    & Deviation of SNs within cluster $C_i$ \\
$\delta_{i}$                    & Energy-based distance coefficient within cluster $C_i$ \\
$s\!_j$                         & $j^{\text{th}}$ SN \\
$s_r^i$                         & $r^{\text{th}}$ SN in the cluster $i$ \\
$\mathbb{S}$                    & Set of SNs \\
$\nu$                           & Percentage of advanced SNs \\
$\psi(j,i)$                     & Distance coefficient of SN $s\!_j$ in a cluster $C_i$ \\
$x_i$, $x\!_j$                  & $x$ coordinate of the SN $s\!_j$ \\
$y_i$, $y\!_j$                  & $y$ coordinate of the centroid $O_j$ \\
$\zeta$                         & Number of remaining SNs in a round \\
$\Phi(j)$                       & Stability function \\ \hline
\end{tabular}}
\caption{Summary of Notations}
%\captionsetup{justification=centering}
\end{table*}

%% file: WSN_MathModel.tex
\section{Mathematical Model}
\label{sec:MathMod}

The model of Adapt-$P$~\cite{suleiman2022cost} formulates the cluster optimality $\kappa_{\text{\tiny{max}}}$ per round required to mitigate the energy consumption of the network to be:
\begin{equation}
\label{equ:Kmax_new}
\begin{split}
   \kappa_{\text{\tiny{max}}} & = \frac{a^2}{2\pi d_{\text{\tiny{opt}}}^2} \\
                              & = \sqrt{\frac{n\epsilon_{{\text{fs}}}}{2\pi \epsilon_{{\text{mp}}}}} ~ \frac{a}{d_{\text{bs}}^2}
\end{split}
\end{equation}
in which $d_{\text{\tiny{opt}}}$ is derived to be the optimal distance between a cluster-head and its cluster-members required to reduce the network's energy consumption, whereas $d_{\text{bs}}$ is the distance from a particular cluster-head to the BS.

The \nolinebreak{$P\!\!\!_{\text{c}} \kappa_{\text{\tiny{max}}}$\small{-}means\small{++}} clustering algorithm starts by utilizing the optimal clustering $\kappa_{\text{\tiny{max}}}$ to define the initial, optimal number of clusters to employ in \nolinebreak{$\kappa$\small{-}means\small{++}} algorithm as follows:
\begin{equation}
\label{equ:Ci}
   \mathbb{C} = \{1, 2, 3, ... , i, ..., \kappa_{\text{\tiny{max}}}\}
\end{equation}
\begin{equation}
\label{equ:Hi}
   \mathbb{H} = \{1, 2, 3, ... , i, ..., \kappa_{\text{\tiny{max}}}\}
\end{equation}
where $C_i$ denotes a cluster with ID $i$ from the set of clusters $\mathbb{C}$ supported with a cluster-head $h_i$ from the set of cluster-heads $\mathbb{H}$, in which $N_i$ represents the number of SNs within the cluster $i$ (including the cluster-head $h_i$) and $s_r^i$ represents the $r^{\text{th}}$ SN in the cluster $i$ where $\nolinebreak{r\!\in\![1,N_i]}$.

\subsection{Cluster-Head Selection Probability $P\!\!\!_{\text{c}}$}

The LEACH algorithm adopts a fixed probability to select a cluster-head by utilizing the $T\!(s)$ function~\cite{Heinzelman}. A cluster-head $h_i$ in Adapt-$P$~\cite{suleiman2022cost} is identified based on the probability $P\!\!\!_{\text{\tiny{adp}}}$ of a SN to become a cluster-head that has been actively adaptive to the state of WSN, characterized by the optimal number of cluster-heads $\kappa_{\text{\tiny{max}}}$ permitted per round and the number of alive SNs $\zeta$ remained in the network as follows:
\begin{equation}
\label{equ:Padp}
   P\!\!\!_{\text{\tiny{adp}}} = \frac{\kappa_{\text{\tiny{max}}}}{\zeta}
\end{equation}

To formulate centroid locations of clusters and accordingly specify SNs to act as cluster-heads, a centroid $O_i$ is selected uniformly at random from the set $\mathbb{S}$ of SNs. Initially, a cluster $C_i$ is composed of the set $\mathbb{S}$ of SNs as follows:
\begin{equation}
\label{equ:Cinitial}
   C\!_i = \{\mathbb{S}\} = \{s_1, s_2, s_3, ..., s\!_j, ..., s_n\}
\end{equation}

The SN $s\!_j$ nearest to the centroid $O_i$ is selected to act as a cluster-head $h_i$ for the initial cluster $C_i$, using the Euclidian formula as follows:
\begin{equation}
\label{equ:EucDO}
   (s\!_j \equiv h_i) = \text{argmin} \parallel D(s\!_j, O_i)\parallel^2\text{} = \sqrt{(x\!_j - x_i)^2 + (y\!_j - y_i)^2},~~~~~~~\forall j\!\!\in\!\![1, n]
\end{equation}
where $[(x\!_j, y\!_j),(x_i, y_i)]$ are $x$ and $y$ coordinates of the SN $s\!_j$ and the centroid $O_i$, respectively.

After which, a distance $D(s\!_j, h_i)$ is computed from each SN $s\!_j$ within $\mathbb{C}$ to the specified cluster-head $h_i$ using the Euclidian formula as follows:
\begin{equation}
\label{equ:EucDS}
   D(s\!_j, h_i) = \sqrt{(x\!_j - x_i)^2 + (y\!_j - y_i)^2},~~~~~~~\forall j\!\!\in\!\![1, n]
\end{equation}
where $[(x\!_j, y\!_j),(x_i, y_i)]$ are $x$ and $y$ coordinates of the SN $s\!_j$ and the cluster-head $h_i$, respectively. The cluster-head $h_i$ is defined to be the closest to the centroid of the cluster $C_i$. Then, another center $h_{i^*}$ is chosen with a probability proportional to $D(s\!_j, h_{i^*})^2$. The procedure is repeated until $\kappa_{max}$ cluster-heads are localized, in which a SN $s\!_j$ would be associated with a cluster-head $h_i$ using the typical \nolinebreak{$\kappa$\small{-}means} algorithm as follows:
\begin{equation}
\label{equ:argKmeans}
   s\!_j \in C_i=\text{argmin} \parallel D(s\!_j, h_i)\parallel^2\text{}\equiv \Big(D(s\!_j, h_i) < D(s\!_j, h_{i^*})\Big)
\end{equation}
where $\nolinebreak{i\neq i^*}$ and $h_i$ belongs to the cluster $C_i$, for $\forall j\!\!\in\!\![1, n]$ and $\forall i^*\!\!\in\!\![1,\kappa_{\text{\tiny{max}}}]$.

The \nolinebreak{$P\!\!\!_{\text{c}} \kappa_{\text{\tiny{max}}}$\small{-}means\small{++}} clustering algorithm formulates that a SN $s\!_j$ joins a cluster-head $h_i$ based on not only the distance $D(s\!_j,h_i)$ between them, but also the residual energy $\varepsilon_i$ of the cluster-head $h_i$. For this purpose, a stability function $\Phi(j)$ is developed based on energy and distance factors of SNs to decide on the cluster-head $h_i$ to which the SN $s\!_j$ should belong.
The \nolinebreak{$P\!\!\!_{\text{c}} \kappa_{\text{\tiny{max}}}$\small{-}means\small{++}} utilizes \nolinebreak{$\kappa$\small{-}means\small{++}} to initialize the first cluster-head and establishes subsequent $\kappa_{\text{\tiny{max}}}$ clusters with a cluster-head selection probability $P\!\!\!_{\text{c}}$ formulated as follows:
\begin{equation}
\label{equ:Pch}
   P\!\!_{\text{c}} = P\!\!\!_{\text{\tiny{adp}}} \times P\!\!\!_{\eta(j,i)} \times P\!\!\!_{\psi(j,i)}, \;\;\;\; \forall i\!\!\in\!\![1,\kappa_{\text{\tiny{max}}}]
\end{equation}
in which $P\!\!\!_{\eta(j,i)}$ is an energy-based quality probability and $P\!\!\!_{\psi(j,i)}$ is a distance-based quality probability explained in the following sections.

\subsection{Energy-based Probability $P\!\!\!_{\eta(j,i)}$}

The $P\!\!\!_{\eta(j,i)}$ is developed to measure the quality of a cluster-head $h_i$ (a SN $s\!_j$ in a cluster $C_i$) according to its residual (remaining) energy-level $\varepsilon_{\text{\tiny{res}}}^{(j,i)}$, initial energy-level $\varepsilon_{\text{\tiny{init}}}^{(j,i)}$ at the time of deployment, and the average energy-level $\varepsilon_{\text{\tiny{avg}}}^i$ of the cluster $C_i$ to which the SN $s\!_j$ belongs. As such, the average residual energy $\varepsilon_{\text{avg}}^i$ of the cluster $C_i$ in a particular epoch is calculated as follows:
\begin{equation}
\label{equ:Eavg}
    \varepsilon_{\text{\tiny{avg}}}^i = \frac{1}{N_i} \sum_{r=1}^{N_i} \varepsilon_{\text{\tiny{res}}}^{(r,i)}    \;\;\;\;\;\; \forall r\in[1, N_i]
\end{equation}  

Then, a global quality probability $P\!_{\xi(j,i)}$ of a SN acting as a cluster-head is formulated by considering SN's residual energy $\varepsilon_{\text{\tiny{res}}}^{(j,i)}$ with respect to the average energy $\varepsilon_{\text{\tiny{avg}}}^i$ of the cluster $C_i$ to which the SN $s\!_j$ belongs to, as follows:
\begin{equation}
\label{equ:Eres_Eavg}
    P\!_{\xi(j,i)} =
    \begin{cases}
         \frac{\varepsilon_{\text{\tiny{res}}}^{(j,i)}}{\varepsilon_{\text{\tiny{avg}}}^i} \;\;\;\;\;\;\;\; , \;\;\; \text{if}~\varepsilon_{\text{\tiny{res}}}^{(j,i)}<{\varepsilon_{\text{\tiny{avg}}}^i} \\
         \;  1                   \;\;\;\; \;\;\;\;  \;\;\;\; ,               \;\;\; \text{otherwise}
    \end{cases}
\end{equation}
in which the global quality probability $P\!\!_{\xi(j,i)}$ is set to 1 if $\varepsilon_{\text{\tiny{res}}}^{(j,i)}$ is higher than ${\varepsilon_{\text{\tiny{avg}}}^i}$ so that an over-estimation is avoided. Such setting qualifies a SN to act as a cluster-head by increasing its election probability as long as its residual energy $\varepsilon_{\text{\tiny{res}}}^{(j,i)}$ is within a particular limit.

In addition, a local quality probability $P\!\!\!_{\varrho(j,i)}$ of a SN to act as a cluster-head is calculated by a typical consideration of its residual energy $\varepsilon_{\text{\tiny{res}}}^{(j,i)}$ with respect to its initial energy $\varepsilon_{\text{\tiny{init}}}^{(j,i)}$ at the time of deployment as follows:
\begin{equation}
\label{equ:Eres_E0}
    P\!\!\!_{\varrho(j,i)} = \frac{\varepsilon_{\text{\tiny{res}}}^{(j,i)}}{\varepsilon_{\text{\tiny{init}}}^{(j,i)}}
\end{equation}

After that, the energy-based probability $P\!\!\!_{\eta(j,i)}$ evaluates the energy effectiveness of a SN $s\!_j$ to act as a cluster-head $h_i$ as follows:
\begin{equation}
\label{equ:f_energy}
   P\!\!\!_{\eta(j,i)} = P\!\!\!_{\varrho(j,i)} \times P\!_{\xi(j,i)}
\end{equation}
which assesses the quality value of the SN $s_j$ to which other SNs would potentially consider to join if it acts as a cluster-head, and accordingly the goal is to maximize $P\!\!_{\eta(j,i)}$ as follows:
\begin{equation}
\label{equ:Peta_max}
   \text{maximize} \; \Big( P\!\!_{\eta(j,i)} \Big) \equiv \text{maximize} \; \Big(P\!\!_{\varrho(j,i)} \times P\!_{\xi(j,i)}\Big)
\end{equation}

\subsection{Distance-based Probability $P\!\!\!_{\psi(j,i)}$}

The $P\!\!\!_{\psi(j,i)}$ is developed to evaluate and measure the positioning $D(s\!_j, h_i)$ of a SN $s\!_j$ toward a cluster-head, relevant to the distance optimality $d_{\text{\tiny{opt}}}$ formalized in Adapt-$P$~\cite{suleiman2022cost} to reduce energy consumption derived to be as follows: \begin{equation}
\label{equ:dopt_new}
    d_{\text{\tiny{opt}}} = \sqrt[4]{\frac{\epsilon_{{\text{mp}}} a^2}{2\pi n\epsilon_{\text{fs}}}} d_{\text{bs}}
\end{equation}

An optimal-oriented distance quality probability $P\!\!_{\gamma(j,i)}$ of a SN $s\!_j$ is formulated based on distance optimality $d_{\text{\tiny{opt}}}$ with respect to a cluster-head $h_i$ that would potentially join as follows:
\begin{equation}
\label{equ:f_1}
    P\!\!_{\gamma(j,i)} =
    \begin{cases}
         \frac{d_{\text{\tiny{opt}}}}{D(s\!_j, h_i)}        \;\;\;\;\;\;\;\; ,  \;\;\; \text{if}~D(s\!_j, h_i)>d_{\text{\tiny{opt}}}\\
         \;\;\;\;\;\; 1 \;\;\;\;\;\;\;\;\;\;\; ,               \;\;\; \text{otherwise}
    \end{cases}
\end{equation}
in which the quality ratio $P\!\!_{\gamma(j,i)}$ is set to 1 if $D(s\!_j, h_i)$ is less than $d_{\text{\tiny{opt}}}$ because the SN $s\!_j$ is within the range of optimal distance $d_{\text{\tiny{opt}}}$ derived to achieve energy-efficient clustering. Such organization increases the probability of a SN $s\!_j$ to join the cluster-head $h_i$ as long as the SN $s\!_j$ is within the optimal range of the cluster formulation.

Furthermore, a typical deviation-based distance quality probability $P\!\!\!_{\sigma(j,i)}$ is derived to measure how a SN $s\!_j$ is distant from a cluster-head $h_i$ based on $d_{\text{\tiny{opt}}}$. First, the deviation $\Delta_{i}$ of all SNs within a cluster $C_i$ is calculated with respect to $d_{\text{\tiny{opt}}}$ as follows:
\begin{equation}
\label{equ:Delta_ji}
   \Delta_{i} = \sum_{j=1}^{N_i} \big( D(s\!_j, h_i) - d_{\text{\tiny{opt}}} \big)^2
\end{equation}
Then, an energy-based distance coefficient $\delta_{i}$ is computed to measure the deviation $\Delta_{i}$ of a SN $s\!_j$ relevant to all SNs $N_i$ within the cluster $C_i$ as follows:
\begin{equation}
\label{equ:delta_coeff}
   \delta_{i} = \sqrt{\frac{\Delta_{i}}{N_i}}
\end{equation}
Accordingly, the deviation-based distance quality probability $P\!\!\!_{\sigma(j,i)}$ of a SN $s\!_j$ in a cluster $C_i$ with $N_i$ SNs associated with its cluster-head $h_i$ is calculated as follows:
\begin{equation}
\label{equ:f_2}
    P\!\!\!_{\sigma(j,i)} =
    \begin{cases}
         \frac{D^2(s\!_j, h_i)}{\sum_{j=1}^{N_i}D^2(s\!_j, h_i)}     \;\;\;\;\; ,    \;\;\; \text{if}~D(s\!_j, h_i)>\delta_{i}\\
         \;\;\;\;\;\; 1 \;\;\;\;\;\;\;\;\;\;\;\;\;\;\;\;\; ,              \;\;\;\;\;\; \text{otherwise}
    \end{cases}
\end{equation}
in which the quality ratio $P\!\!\!_{\sigma(j,i)}$ is set to 1 if energy-based distance coefficient $\delta_{i}$ is higher than $D(s\!_j, h_i)$ because the SN $s\!_j$ is within the range of its $\delta_{i}$ derived previously.

After that, the distance-based probability $P\!\!\!_{\psi(j,i)}$ is formulated as follows:
\begin{equation}
\label{equ:f_distance}
   P\!\!\!_{\psi(j,i)} = P\!\!_{\gamma(j,i)} \times P\!\!\!_{\sigma(j,i)}
\end{equation}
which assesses the efficacy value of the distance of a SN $s_j$ toward a cluster-head $h_i$ that would consider to join, and accordingly the goal is to maximize the $P\!\!\!_{\psi(j,i)}$ as follows.
\begin{equation}
\label{equ:Ppsi_max}
   \text{maximize} \; \Big( P\!\!\!_{\psi(j,i)} \Big) \equiv \text{maximize} \; \Big(P\!\!_{\gamma(j,i)} \times P\!\!\!_{\sigma(j,i)}\Big)
\end{equation}

\subsection{Formalizing $\Phi(j)$}

The objective is to maximize the stability function $\Phi(j)$ for a SN $s\!_j$ joining a cluster-head $h_i$ so that energy-efficient clusters are produced, as follows:
\begin{equation}
\label{equ:f_total}
   \text{maximize} \; \Big( \Phi(j) \Big) \equiv \text{maximize} \; \Big(\eta(j,i) \times \psi(j,i)\Big)
\end{equation}
in which the $\Phi(j)$ is computed based on the energy quality factor $\eta(j,i)$ of a potential cluster-head $h_i$ and the distance quality factor $\psi(j,i)$ of the SN $s_j$ entailed with the cluster-head $h_i$, in which the cluster-head election probability $P\!\!_{\text{c}}$ is maximized:
\begin{equation}
\label{equ:f_total}
   \text{maximize} \; \Big( P\!\!_{\text{c}} \Big) \equiv \text{maximize} \; \Big(P\!\!\!_{\text{\tiny{adp}}} \times P\!\!\!_{\eta(j,i)} \times P\!\!\!_{\psi(j,i)}\Big)
\end{equation}

%% file: WSN_Evaluation.tex
\section{Evaluation}
\label{sec:eval}

The \nolinebreak{$P\!\!\!_{\text{c}} \kappa_{\text{\tiny{max}}}$\small{-}means\small{++}} clustering is employed in the SEP algorithm, in which the probability $P\!\!\!_{\text{c}}$ of selecting a cluster-head $h_i$ in a cluster $C_i$ is developed based on the adaptive probability $P\!\!\!_{\text{\tiny{adp}}}$, energy-based probability $P\!\!\!_\eta$, and distance-based probability $P\!\!\!_\psi$ to accurately formulate clusters from the set $\mathbb{S}$ of SNs. The design is also developed based on the LEACH algorithm to compare the performance of such clustering schemes together. MATLAB is the platform used to simulate and demonstrate the efficacy of such algorithms.

\subsection{Analysis of Remaining Energy}

The \nolinebreak{$P\!\!\!_c \kappa_{\text{\tiny{max}}}$\small{-}means\small{++}}, \nolinebreak{$P\!\!\!_\eta \varepsilon \kappa_{\text{\tiny{max}}}$\small{-}means\small{++}}, and \nolinebreak{$P\!\!\!_\psi d_{\text{\tiny{opt}}} \kappa_{\text{\tiny{max}}}$\small{-}means\small{++}} algorithms are proposed and compared by analyzing the remaining energy in SNs. The performance is analyzed by measuring the impacts of energy coefficient $\eta(j,i)$, distance coefficient $\psi(j,i)$, and stability function $\Phi(j)$ on the residual energy of SNs.

\subsubsection{Impact of Energy Coefficient $\eta(j,i)$}

The probability $P\!\!\!_\eta$ of electing a cluster-head $h_i$, the energy coefficient $\varepsilon$ of the SN, and the optimal number $\kappa_{\text{\tiny{max}}}$ of cluster-heads permitted per round are utilized to develop energy-based algorithms that are \nolinebreak{$P\!\!\!_\eta \varepsilon \kappa_{\text{\tiny{max}}}$\small{-}means\small{++}}, \nolinebreak{$P\!\!\!_\eta \varepsilon \kappa_{\text{\tiny{max}}}$\small{-}\footnotesize{SEP}}, and \nolinebreak{$P\!\!\!_\eta \varepsilon \kappa_{\text{\tiny{max}}}$\small{-}\footnotesize{LEACH}}. Their development is based on \nolinebreak{$\kappa$\small{-}means\small{++}}, \nolinebreak{\footnotesize{SEP}}, and \nolinebreak{\footnotesize{LEACH}} algorithms.
\begin{figure}[!h]
\centering
\captionsetup{justification=centering}
	  \includegraphics[width=\textwidth]{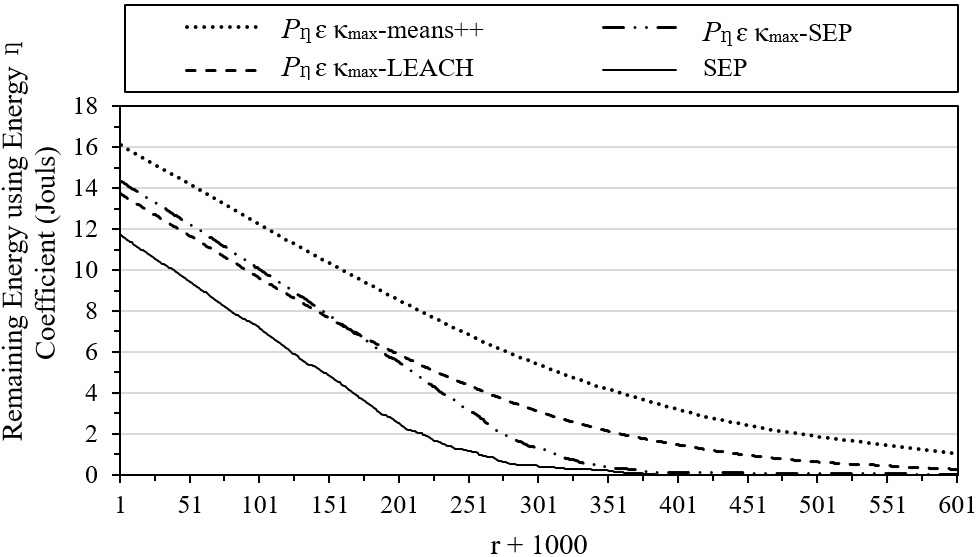}
	  \caption{\small{Impact of Energy Coefficient $\eta(j,i)$ on Remaining Energy}}
      \label{fig:fig1}
\end{figure}

Figure~\ref{fig:fig1} compares the residual energy of the set $\mathbb{S}$ of SNs for such energy-based algorithms. It is shown that the performance of \nolinebreak{$P\!\!\!_\eta \varepsilon \kappa_{\text{\tiny{max}}}$\small{-}means\small{++}} algorithm increases the remaining energy of SNs as compared to the rest of proposed algorithms. The reason is that the cluster-head selection probability is evolved based on the energy-based probabilistic value $P\!\!\!_{\eta(j,i)}$ of a potential cluster-head $h_i$ and the adaptive probability $P\!\!\!_{\text{\tiny{adp}}}$ of selecting a cluster-head. The $P\!\!_{\eta(j,i)}$ effectively boosts a SN $s\!_j$ to become a cluster-head by maximizing SN's global $P_{\xi(j,i)}$ and local $P\!\!\!_{\varrho(j,i)}$ probabilistic ratios. The global $P_{\xi(j,i)}$ qualifies a SN $s\!_j$ to become a cluster-head in a cluster $C_i$ by increasing its election probability provided that its residual energy $\varepsilon_{\text{\tiny{res}}}^{(j,i)}$ is higher than the average $\varepsilon_{\text{\tiny{avg}}}^i$ residual energy of SNs of the cluster $C_i$. On the other side, the local $P\!\!\!_{\varrho(j,i)}$ increases the probability of a SN $s\!_j$ to become a cluster-head if its residual energy $\varepsilon_{\text{\tiny{res}}}^{(j,i)}$ is close to its initial energy $\varepsilon_{\text{\tiny{init}}}^{(j,i)}$. It is asserted that both probabilistic ratios tend to set and thus increase the probability of a SN to become a cluster-head as long as the quality value of SN's energy is above the specified threshold.

In addition, the \nolinebreak{$P\!\!\!_\eta \varepsilon \kappa_{\text{\tiny{max}}}$\small{-}means\small{++}} typically saves and facilitates the time to classify a large WSN. Although the \nolinebreak{$\kappa$\small{-}means\small{++}} is computationally expensive in which centroids in the first rounds are initially located in various places in the field, the convergence running-time required to construct optimal clusters is minimal and the intra-cluster variance is mitigated. The optimal clustering $\kappa_{\text{\tiny{max}}}$ calculated in advance supports the election process by avoiding any over- or under-estimation to the number and size of clusters to initiate. Such optimality through $\kappa_{\text{\tiny{max}}}$ within \nolinebreak{$\kappa$\small{-}means\small{++}} mitigates any extra communication between SNs so that their clustering layouts are uniformly formulated, and accordingly energy of SNs would be saved. Incorporating the probability $P\!\!\!_\eta$ provides a high priority to a SN whose energy is above pre-specified energy thresholds in which the probability is set to 1 as long as the residual energy $\varepsilon_{\text{\tiny{res}}}^{(j,i)}$ of the SN $s\!_j$ is within its specified acceptable margins. Such election enforcements potentially promote the likelihood of selecting the SN to act as a cluster-head.

The \nolinebreak{$P\!\!\!_\eta \varepsilon \kappa_{\text{\tiny{max}}}$-\footnotesize{SEP}} outperforms the \nolinebreak{$P\!\!\!_\eta \varepsilon \kappa_{\text{\tiny{max}}}$-\footnotesize{LEACH}} at the beginning by preserving more energy in SNs $\zeta$ remained in the network. The \nolinebreak{\footnotesize{SEP}} protocol actually accounts for SNs' heterogeneity, in which \nolinebreak{\footnotesize{SEP}} considers advanced and normal SNs with different initial energy levels whereas the formulation of \nolinebreak{\footnotesize{LEACH}} protocol does not generally take into consideration such energy variations. The \nolinebreak{$P\!\!\!_\eta \varepsilon \kappa_{\text{\tiny{max}}}$-\footnotesize{LEACH}} algorithm shows a little enhancement at the end of rounds, however impacts of $P\!\!\!_\eta$ and $\kappa_{\text{\tiny{max}}}$ coefficients along with the energy factor $\varepsilon$ typically preserve the performance of \nolinebreak{\footnotesize{SEP}}-based and \nolinebreak{\footnotesize{LEACH}}-based algorithms to generally make them perform similarly.
Furthermore, the average energy saving per round for the proposed algorithms in the specified range shown in Figure~\ref{fig:fig1} reflects such findings. For instance, the \nolinebreak{$P\!\!\!_\eta \varepsilon \kappa_{\text{\tiny{max}}}$\small{-}means\small{++}} algorithm preserves the energy of SNs by showing an amount of $6.58$ \nolinebreak{Jouls/round}. The \nolinebreak{$P\!\!\!_\eta \varepsilon \kappa_{\text{\tiny{max}}}$-\footnotesize{SEP}} and \nolinebreak{$P\!\!\!_\eta \varepsilon \kappa_{\text{\tiny{max}}}$-\footnotesize{LEACH}} algorithms have approximately $3.97$ and $4.56$ \nolinebreak{Jouls/round}, respectively. The \nolinebreak{\footnotesize{SEP}} algorithm has the lowest average energy, showing an amount of $2.6$ \nolinebreak{Jouls/round}. The \nolinebreak{$P\!\!\!_\eta \varepsilon \kappa_{\text{\tiny{max}}}$\small{-}means\small{++}} has overall improved the performance and thus network lifetime, as well as efficiently covers the networking space.

\subsubsection{Impact of Distance Coefficient $\psi(j,i)$}

The optimal distance $d_{\text{\tiny{opt}}}$ is utilized to propose \nolinebreak{$P\!\!\!_\psi d_{\text{\tiny{opt}}} \kappa_{\text{\tiny{max}}}$\small{-}means\small{++}}, \nolinebreak{$P\!\!\!_\psi d_{\text{\tiny{opt}}} \kappa_{\text{\tiny{max}}}$\small{-}\footnotesize{SEP}}, and \nolinebreak{$P\!\!\!_\psi d_{\text{\tiny{opt}}} \kappa_{\text{\tiny{max}}}$\small{-}\footnotesize{LEACH}} algorithms. The performance of such distance-based clustering schemes is evaluated using the distance-based probability $P\!\!\!_{\psi(j,i)}$ and distance factor $\psi(j,i)$.
\begin{figure}[!h]
\centering
\captionsetup{justification=centering}
	  \includegraphics[width=\textwidth]{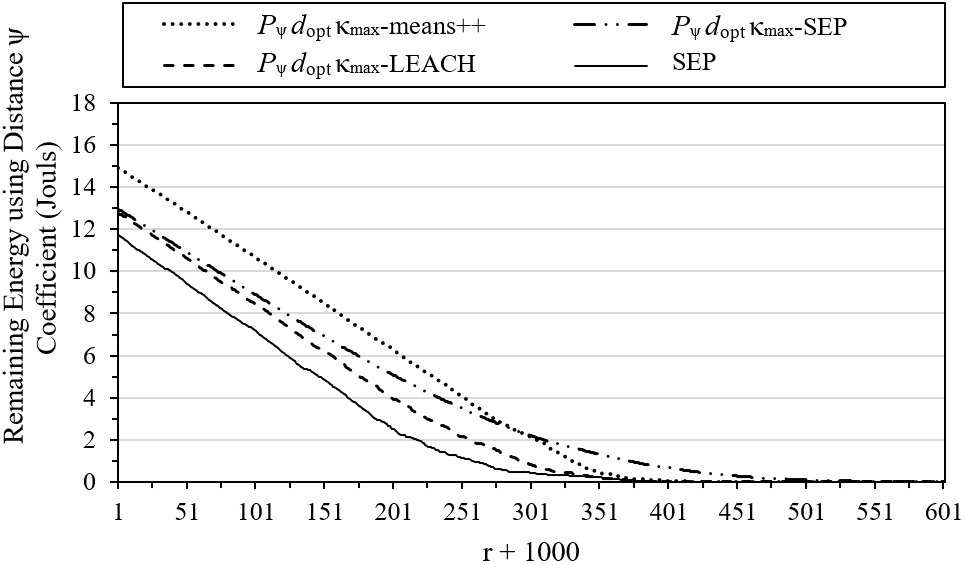}
	  \caption{\small{Impact of Distance Coefficient $\psi(j,i)$ on Remaining Energy}}
      \label{fig:fig2}
\end{figure}

The residual energy of the set $\mathbb{S}$ of SNs based on the distance-based algorithms is shown in Figure~\ref{fig:fig2}. The \nolinebreak{$P\!\!\!_\psi d_{\text{\tiny{opt}}} \kappa_{\text{\tiny{max}}}$\small{-}means\small{++}} outperforms the rest of clustering schemes, in which the distance-based probabilistic value $P\!\!\!_{\psi(j,i)}$ adapts the cluster-head selection probability based on its distance factor $\psi(j,i)$. In such formulations, the distance quality probability $P\!\!_{\gamma(j,i)}$ of a SN $s\!_j$ is increased provided that SN's position $D(s\!_j, h_i)$ is localized within an acceptable margin of the optimal distance $d_{\text{\tiny{opt}}}$ of the cluster $C_i$, in which the quality ratio of the SN $s\!_j$ is thus increased. In contrast, the probability $P\!\!_{\gamma(j,i)}$ decreases with the increment of the distance $D(s\!_j, h_i)$ from the cluster-head $h_i$.
In addition, the deviation $\Delta_{j,i}$ of SNs within a cluster $C_i$ proportionally impacts the distance coefficient $\delta_{j,i}$ based on $d_{\text{\tiny{opt}}}$ to formulate energy-efficient clusters. The deviation-based distance probabilistic ratio $P\!\!\!_{\sigma(j,i)}$ is affected by such formations and increased provided that the SN $s\!_j$ is localized in a place close to its distance coefficient $\delta_{j,i}$, which overall increases the SN's distance-based probability $P\!\!\!_{\psi(j,i)}$.

Furthermore, Figure~\ref{fig:fig2} demonstrates that the efficacy of the \nolinebreak{$P\!\!\!_\psi d_{\text{\tiny{opt}}} \kappa_{\text{\tiny{max}}}$\small{-}means\small{++}} algorithm in clustering SNs in an efficient manner. The distance optimality $d_{\text{\tiny{opt}}}$ controls the distances within clusters, in which clusters are accordingly formed with respect to $\kappa_{\text{\tiny{max}}}$ so that extra cluster formations are avoided. The election probability also promotes SNs to act as cluster-heads by setting the distance-based probability $P\!\!\!_{\psi(j,i)}$ to $1$ for SNs that are distant by at most $d_{\text{\tiny{opt}}}$ from their corresponding cluster-heads. The \nolinebreak{$P\!\!\!_\psi d_{\text{\tiny{opt}}} \kappa_{\text{\tiny{max}}}$-\footnotesize{SEP}} outperforms the \nolinebreak{$P\!\!\!_\psi d_{\text{\tiny{opt}}} \kappa_{\text{\tiny{max}}}$-\footnotesize{LEACH}}, which in turn surpasses the \nolinebreak{\footnotesize{SEP}} algorithm. The average amount of energy saving in the specified range represented in Figure~\ref{fig:fig2} for \nolinebreak{$P\!\!\!_\psi d_{\text{\tiny{opt}}} \kappa_{\text{\tiny{max}}}$\small{-}means\small{++}} reaches around $4.34$ \nolinebreak{Jouls/round}, whereas \nolinebreak{$P\!\!\!_\psi d_{\text{\tiny{opt}}} \kappa_{\text{\tiny{max}}}$-\footnotesize{SEP}} and \nolinebreak{$P\!\!\!_\psi d_{\text{\tiny{opt}}} \kappa_{\text{\tiny{max}}}$-\footnotesize{LEACH}} algorithms show respectively $3.84$ and $3.21$ \nolinebreak{Jouls/round}. The \nolinebreak{\footnotesize{SEP}} algorithm as discussed formerly depicts $2.6$ \nolinebreak{Jouls/round}. Such findings demonstrate the effective performance proven by \nolinebreak{$P\!\!\!_\psi d_{\text{\tiny{opt}}} \kappa_{\text{\tiny{max}}}$\small{-}means\small{++}} as compared to the proposed distance-based algorithms, as well as corroborates the performance of energy-based algorithms developed previously in which the \nolinebreak{$P\!\!\!_\eta \varepsilon \kappa_{\text{\tiny{max}}}$\small{-}means\small{++}} outperforms the rest of proposed algorithms.

\subsubsection{Impact of Stability Function $\Phi(j)$}

\begin{figure}[!t]
\centering
\captionsetup{justification=centering}
	  \includegraphics[width=\textwidth]{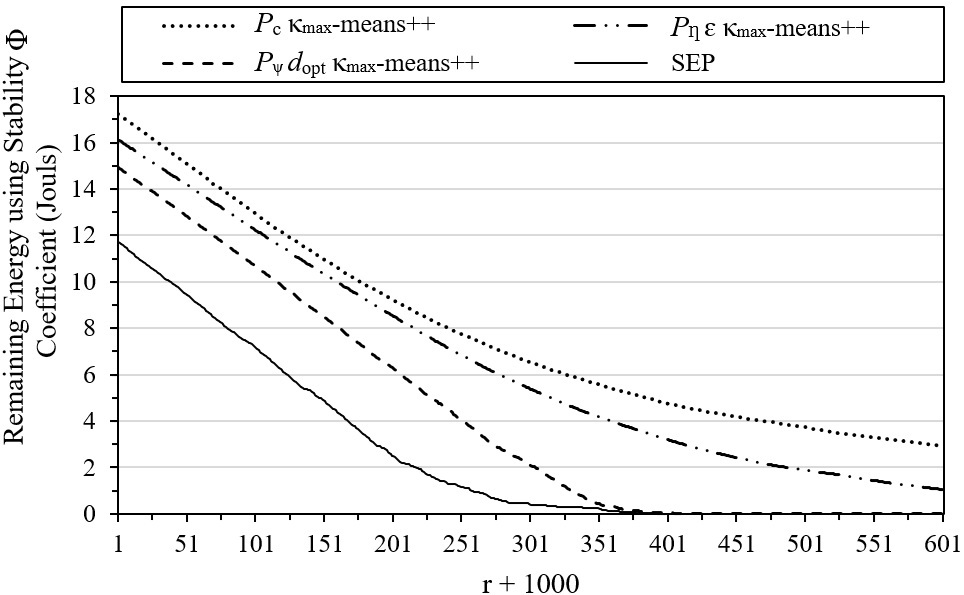}
	  \caption{\small{Impact of Stability Function $\Phi(j)$ on Remaining Energy}}
      \label{fig:fig3}
\end{figure}

The performance effect of \nolinebreak{$P\!\!\!_{\text{c}} \kappa_{\text{\tiny{max}}}$\small{-}means\small{++}} is evaluated with \nolinebreak{$P\!\!\!_\eta \varepsilon \kappa_{\text{\tiny{max}}}$\small{-}means\small{++}} and \nolinebreak{$P\!\!\!_\psi d_{\text{\tiny{opt}}} \kappa_{\text{\tiny{max}}}$\small{-}means\small{++}} clustering schemes. Figure~\ref{fig:fig3} demonstrates the effectiveness of incorporating the election probability $P\!\!\!_{\text{c}}$ of a cluster-head $h_i$ in which \nolinebreak{$P\!\!\!_{\text{c}} \kappa_{\text{\tiny{max}}}$\small{-}means\small{++}} outperforms the rest of proposed algorithms. The cluster-head election probability $P\!\!\!_{\text{c}}$ is evolved by employing the energy-based $P\!\!\!_{\eta(j,i)}$ and distance-based $P\!\!\!_{\psi(j,i)}$ quality probabilities, in which the stability function $\Phi(j)$ is optimized by maximizing the energy quality factor $\eta(j,i)$ and distance quality factor $\psi(j,i)$ of the SN $s\!_j$.

The adaptive election probability $P\!\!\!_{\text{\tiny{adp}}}$ within $P\!\!\!_{\text{c}}$ is dependent on SN states as a whole, in which the selection probability of a SN $s\!_j$ is evolved along with the remaining number $\zeta$ of alive SNs in the field and the optimal clustering $\kappa_{\text{\tiny{max}}}$ formulated in a round according to SN distributions and states. The derivation and optimization of $\kappa_{\text{\tiny{max}}}$ within $P\!\!\!_{\text{\tiny{adp}}}$ and hence within $P\!\!\!_{\text{c}}$ limits the number of cluster formations in a particular epoch based on the optimal distance $d_{\text{\tiny{opt}}}$ such that the residual energy of the network is maximized, which in turn formulates clusters that uniformly cover the networking space and efficiently preserve the residual energy of SNs. The average amount of energy saving in the specified range in Figure~\ref{fig:fig3} for \nolinebreak{$P\!\!\!_{\text{c}} \kappa_{\text{\tiny{max}}}$\small{-}means\small{++}} is around $7.82$ \nolinebreak{Jouls/round}, as compared to \nolinebreak{$P\!\!\!_\eta \varepsilon \kappa_{\text{\tiny{max}}}$\small{-}means\small{++}} and \nolinebreak{$P\!\!\!_\psi d_{\text{\tiny{opt}}} \kappa_{\text{\tiny{max}}}$\small{-}means\small{++}} algorithms that respectively preserve an average energy of $6.58$ \nolinebreak{Jouls/round} and $4.34$ \nolinebreak{Jouls/round}.

\subsection{Analysis of Remaining SNs $\zeta$}

The performance of the proposed clustering algorithms based on the remaining number $\zeta$ of SNs depicts the likelihood of dead SNs in the networking field. The election probability selected to manage the distribution and number of remained SNs in the network impacts the value of $\zeta$. The following analyzes the effect of the energy-based probability $P\!\!\!_{\eta(j,i)}$ driven by the energy quality factor $\eta(j,i)$, the distance-based probability $P\!\!\!_{\psi(j,i)}$ driven by the distance quality factor $\psi(j,i)$, and the election probability $P\!\!\!_{\text{c}}$ driven by the stability function $\Phi(j)$.

\subsubsection{Analysis of $\zeta$ Driven by Energy-based Election Probability $P\!\!\!_{\eta(j,i)}$}

\begin{figure}[!t]
\centering
\captionsetup{justification=centering}
	  \includegraphics[width=\textwidth]{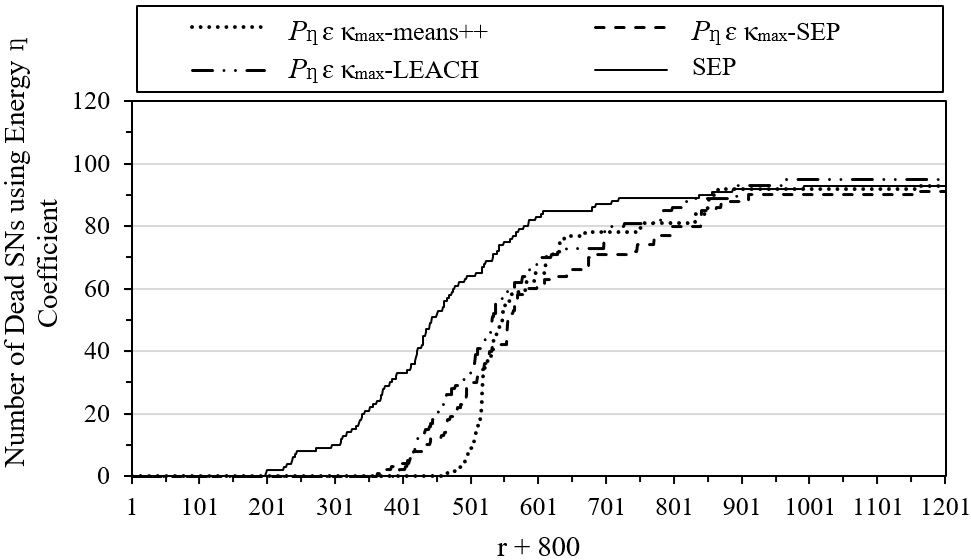}
	  \caption{\small{Analysis of $\zeta$ Driven by the Probability $P\!\!\!_{\eta(j,i)}$}}
      \label{fig:fig4}
\end{figure}

The percentage of remained $\zeta$ SNs in the network is analyzed as a result of employing the probability $P\!\!\!_{\eta(j,i)}$ in the formulation of clusters. Figure~\ref{fig:fig4} demonstrates the efficacy of \nolinebreak{$P\!\!\!_\eta \varepsilon \kappa_{\text{\tiny{max}}}$\small{-}means\small{++}} clustering algorithm in improving $\zeta$. Employing $P\!\!\!_{\eta(j,i)}$ along with \nolinebreak{$\kappa_{\text{\tiny{max}}}$\small{-}means\small{++}} clustering scheme fosters and stabilizes the process of cluster formation. The election probability $P\!\!\!_{\eta(j,i)}$ of a SN to act as a cluster-head supports such enhancements due to its development based on energy consideration in which the global $P\!\!\!_{\varrho(j,i)}$ and local $P_{\xi(j,i)}$ probabilistic ratios are maximized, and potentially can be set to $1$ provided that the SN $s\!_j$ maintains a residual energy $\varepsilon_{\text{\tiny{res}}}^{(j,i)}$ close to its specified $\varepsilon_{\text{\tiny{init}}}^{(j,i)}$ and $\varepsilon_{\text{\tiny{avg}}}^i$ margins.

The \nolinebreak{$P\!\!\!_\eta \varepsilon \kappa_{\text{\tiny{max}}}$-\footnotesize{SEP}} increases the percentage $\zeta$ of remained SNs more than the \nolinebreak{$P\!\!\!_\eta \varepsilon \kappa_{\text{\tiny{max}}}$-\footnotesize{LEACH}}, however both algorithms are outperformed by \nolinebreak{$P\!\!\!_\eta \varepsilon \kappa_{\text{\tiny{max}}}$\small{-}means\small{++}} due to the incorporation of the energy-based election probability $P\!\!\!_{\eta(j,i)}$ shaped by the optimal clustering $\kappa_{\text{\tiny{max}}}$ within the deployment of \nolinebreak{$\kappa_{\text{\tiny{max}}}$\small{-}means\small{++}} clustering scheme.
In addition, the stability period defined by the time to death of the first SN varies from an algorithm to another. The \nolinebreak{$P\!\!\!_\eta \varepsilon \kappa_{\text{\tiny{max}}}$\small{-}means\small{++}} takes $1260$ rounds to announce the death of the first SN in the field, whereas \nolinebreak{$P\!\!\!_\eta \varepsilon \kappa_{\text{\tiny{max}}}$-\footnotesize{SEP}} and \nolinebreak{$P\!\!\!_\eta \varepsilon \kappa_{\text{\tiny{max}}}$-\footnotesize{LEACH}} algorithms take respectively $1190$ and $1163$ rounds to flag the death of the first SN. The \nolinebreak{\footnotesize{SEP}} algorithm takes $999$ rounds. The \nolinebreak{$P\!\!\!_\eta \varepsilon \kappa_{\text{\tiny{max}}}$\small{-}means\small{++}} algorithm has overall managed the decrement of $\zeta$ to maintain a feasible network coverage and a maximized residual energy in SNs.

\subsubsection{Analysis of $\zeta$ Driven by Distance-based Election Probability $P\!\!\!_{\psi(j,i)}$}

\begin{figure}[!t]
\centering
\captionsetup{justification=centering}
	  \includegraphics[width=\textwidth]{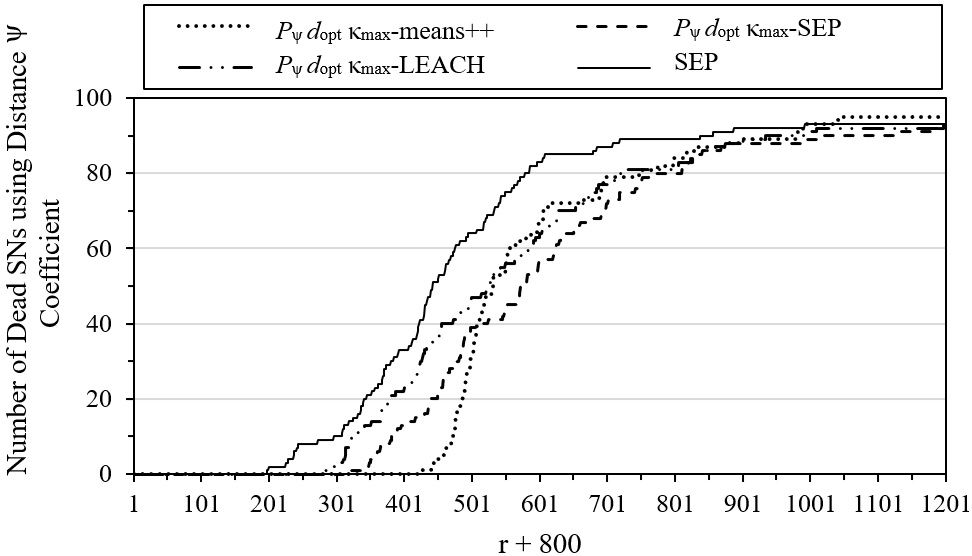}
	  \caption{\small{Analysis of $\zeta$ Driven by the Probability $P\!\!\!_{\psi(j,i)}$}}
      \label{fig:fig5}
\end{figure}

The percentage of remained SNs $\zeta$ in the network is analyzed with respect to the election probability $P\!\!\!_{\psi(j,i)}$ developed based on the optimal distance $d_{\text{\tiny{opt}}}$ derived to formulate energy-efficient clusters. Figure~\ref{fig:fig5} elucidates the efficacy of \nolinebreak{$P\!\!\!_\psi d_{\text{\tiny{opt}}} \kappa_{\text{\tiny{max}}}$\small{-}means\small{++}}, in which the algorithm stabilizes the curve of remained SNs by forming clusters based on distance optimality $d_{\text{\tiny{opt}}}$ such that the residual energy of SNs is maximized. The distance-based election probability $P\!\!\!_{\psi(j,i)}$ integrated with \nolinebreak{$\kappa_{\text{\tiny{max}}}$\small{-}means\small{++}} supports a steady selection of cluster-heads based on the optimal clustering $\kappa_{\text{\tiny{max}}}$ and forms size-balanced clusters based on the optimal distance $d_{\text{\tiny{opt}}}$ for energy efficiency, which in turn increases the percentage of $\zeta$ in the network.
It is shown in Figure~\ref{fig:fig5} that the \nolinebreak{$P\!\!\!_\psi d_{\text{\tiny{opt}}} \kappa_{\text{\tiny{max}}}$\small{-}means\small{++}} outperforms \nolinebreak{$P\!\!\!_\psi d_{\text{\tiny{opt}}} \kappa_{\text{\tiny{max}}}$-\footnotesize{SEP}} and \nolinebreak{$P\!\!\!_\psi d_{\text{\tiny{opt}}} \kappa_{\text{\tiny{max}}}$-\footnotesize{LEACH}} algorithms. The time required to death of the first SN in \nolinebreak{$P\!\!\!_\psi d_{\text{\tiny{opt}}} \kappa_{\text{\tiny{max}}}$\small{-}means\small{++}} is $1225$ rounds, whereas the time to death of the first SN in \nolinebreak{$P\!\!\!_\psi d_{\text{\tiny{opt}}} \kappa_{\text{\tiny{max}}}$-\footnotesize{SEP}} and \nolinebreak{$P\!\!\!_\psi d_{\text{\tiny{opt}}} \kappa_{\text{\tiny{max}}}$-\footnotesize{LEACH}} algorithms is respectively $1076$ and $1125$ rounds.

\subsubsection{Analysis of $\zeta$ Driven by Election Probability $P\!\!\!_{\text{c}}$}

\begin{figure}[!t]
\centering
\captionsetup{justification=centering}
	  \includegraphics[width=\textwidth]{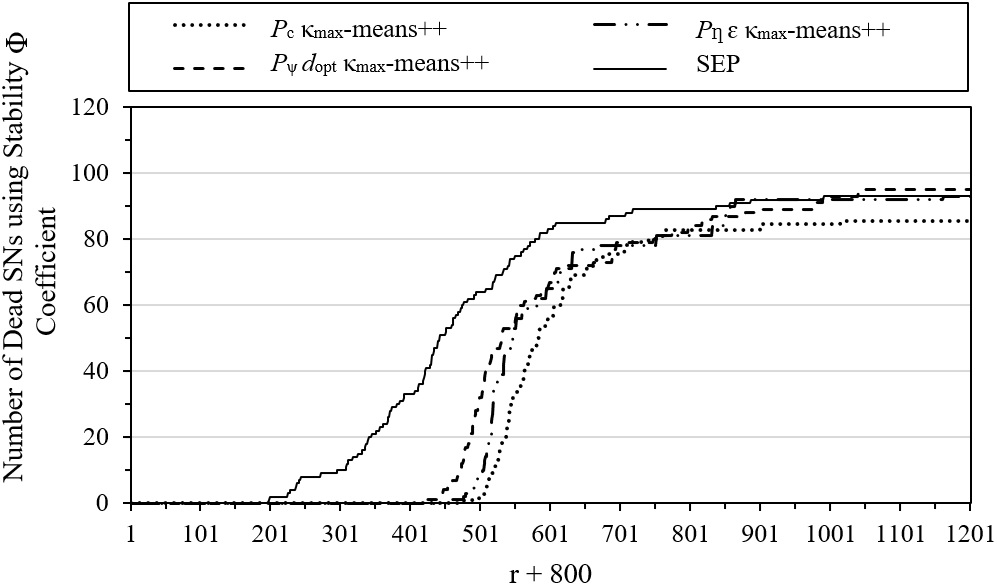}
	  \caption{\small{Analysis of $\zeta$ Driven by the Probability $P\!\!\!_{\text{c}}$}}
      \label{fig:fig6}
\end{figure}

The efficacy of employing the election probability $P\!\!\!_{\text{c}}$ with the \nolinebreak{$\kappa_{\text{\tiny{max}}}$\small{-}means\small{++}} clustering is described in Figure~\ref{fig:fig6}, in which the \nolinebreak{$P\!\!\!_{\text{c}} \kappa_{\text{\tiny{max}}}$\small{-}means\small{++}} is compared with the energy-based \nolinebreak{$P\!\!\!_\eta \varepsilon \kappa_{\text{\tiny{max}}}$\small{-}means\small{++}} and distance-based \nolinebreak{$P\!\!\!_\psi d_{\text{\tiny{opt}}} \kappa_{\text{\tiny{max}}}$\small{-}means\small{++}} clustering schemes. The probability $P\!\!\!_{\text{c}}$ is continuously evolved with variations occurred in network states which as a results promotes the impact of the adaptive probability $P\!\!\!_{\text{\tiny{adp}}}$ controlled by the remaining SNs $\zeta$ and the optimal clustering $\kappa_{\text{\tiny{max}}}$ of the network. The dependency of $P\!\!\!_{\text{c}}$ on energy-based $P\!\!\!_{\eta(j,i)}$ and distance-based $P\!\!\!_{\psi(j,i)}$ quality probabilities manages the formation of clusters by increasing the likelihood of qualified SNs to become cluster-heads.

It is also shown that the energy-based clustering in \nolinebreak{$P\!\!\!_\eta \varepsilon \kappa_{\text{\tiny{max}}}$\small{-}means\small{++}} performs better than the distance-based clustering in \nolinebreak{$P\!\!\!_\psi d_{\text{\tiny{opt}}} \kappa_{\text{\tiny{max}}}$\small{-}means\small{++}}, however, they both at some points behave similarly. The involvement of $\varepsilon$ typically influences the performance of the energy-based algorithm, yet the optimal distance $d_{\text{\tiny{opt}}}$ is also derived based on energy consideration $\varepsilon$ and optimal clustering $\kappa_{\text{\tiny{max}}}$ which makes the distance-based algorithm \nolinebreak{$P\!\!\!_\psi d_{\text{\tiny{opt}}} \kappa_{\text{\tiny{max}}}$\small{-}means\small{++}} exhibit a reasonable performance.
The stability period of energy-based \nolinebreak{$P\!\!\!_\eta \varepsilon \kappa_{\text{\tiny{max}}}$\small{-}means\small{++}} clustering reaches $1260$ rounds, whereas it is $1225$ rounds for the distance-based \nolinebreak{$P\!\!\!_\psi d_{\text{\tiny{opt}}} \kappa_{\text{\tiny{max}}}$\small{-}means\small{++}} clustering. In contrast, the stability period of the \nolinebreak{$P\!\!\!_{\text{c}} \kappa_{\text{\tiny{max}}}$\small{-}means\small{++}} clustering algorithm reaches $1289$ rounds, which performs better than other algorithms.

%% file: WSN_Conc_Future.tex
\section{Conclusion and Future Work}
\label{sec:conc}

It is found that the \nolinebreak{$P\!\!\!_{\text{c}} \kappa_{\text{\tiny{max}}}$\small{-}means\small{++}} clustering algorithm demonstrates an optimized performance in increasing the residual energy of SNs, stability period, and likelihood of remained SNs $\zeta$ in the network. Adaptability in the design has been applied by evolving the cluster-head election probability $P\!\!\!_{\text{c}}$ developed based on energy-based probability $P\!\!\!_{\eta(j,i)}$, distance-based probability $P\!\!\!_{\psi(j,i)}$, and adaptive probability $P\!\!\!_{\text{\tiny{adp}}}$. It is observed that the election probability $P\!\!\!_{\text{c}}$ has boosted the effectiveness of the algorithm to formulate clusters such that energy of SNs is preserved and network lifetime is prolonged.
The involvement of energy $\varepsilon$, optimal clustering $\kappa_{\text{\tiny{max}}}$, and distance optimality $d_{\text{\tiny{opt}}}$ factors derived in Adapt-$P$ along with the \nolinebreak{$k${\small{-}}means{\small{++}}} clustering scheme has led the development of energy-based \nolinebreak{$P\!\!\!_\eta \varepsilon \kappa_{\text{\tiny{max}}}$\small{-}means\small{++}} and distance-based \nolinebreak{$P\!\!\!_\psi d_{\text{\tiny{opt}}} \kappa_{\text{\tiny{max}}}$\small{-}means\small{++}} clustering algorithms, and thus to further enhance the performance in which the optimal clustering $\kappa_{\text{\tiny{max}}}$ has been utilized to formulate adaptive clustering through \nolinebreak{$\kappa_{\text{\tiny{max}}}$\small{-}means\small{++}} algorithm. As a future work, it is noticed that the genetic algorithm has a potential application in such clustering problems in which a chromosome can be formed based on a set of SNs and a fitness function for each chromosome can accordingly be derived based on energy requirements of SNs.

%% file: HusamWSN_6.bbl
\begin{thebibliography}{10}

\bibitem{agrawal22}
V.~Agarwal, S.~Tapaswi, and P.~Chanak.
\newblock Intelligent fault-tolerance data routing scheme for {IoT}-enabled
  {WSNs}.
\newblock {\em IEEE Internet of Things Journal}, 9(17):16332--16342, 2022.

\bibitem{alomari2022systematic}
M.~Alomari, M.~Mahmoud, and M.~Ramli.
\newblock A systematic review on the energy efficiency of dynamic clustering in
  a heterogeneous environment of wireless sensor networks ({WSNs}).
\newblock {\em Electronics}, 11(18):2837, 2022.

\bibitem{ARJUNAN2019}
S.~Arjunan and S.~Pothula.
\newblock A survey on unequal clustering protocols in wireless sensor networks.
\newblock {\em Journal of King Saud University - Computer and Information
  Sciences}, 31(3):304--317, 2019.

\bibitem{kmeans}
D.~Arthur and S.~Vassilvitskii.
\newblock {K-Means++}: The advantages of careful seeding.
\newblock In {\em Proc. of the 18th Annual {ACM-SIAM} Symposium on Discrete
  Algorithms}, page 1027–1035, 2007.

\bibitem{Azzouz2022}
I.~Azzouz, B.~Boussaid, A.~Zouinkhi, and M.~Abdelkrim.
\newblock Energy-aware cluster head selection protocol with balanced fuzzy
  c-mean clustering in {WSN}.
\newblock In {\em Proc. of the International Multi-Conference on Systems,
  Signals \& Devices}, pages 1534--1539, 2022.

\bibitem{Shashi2018}
S.~Bhushan, R.~Pal, and S.~Antoshchuk.
\newblock Energy efficient clustering protocol for heterogeneous wireless
  sensor network: {A} hybrid approach using {GA} and {K}-means.
\newblock In {\em Proc. of the {IEEE} Second International Conference on Data
  Stream Mining \& Processing ({DSMP})}, pages 381--385, 2018.

\bibitem{bidaki2016towards}
M.~Bidaki, R.~Ghaemi, and S.~Tabbakh.
\newblock Towards energy efficient {K-means} based clustering scheme for
  wireless sensor networks.
\newblock {\em International Journal of Grid and Distributed Computing},
  9(7):265--276, 2016.

\bibitem{Ismail20}
I.~Butun, P.~Osterberg, and H.~Song.
\newblock Security of the internet of things: Vulnerabilities, attacks, and
  countermeasures.
\newblock {\em IEEE Communications Surveys \& Tutorials}, 22(1):616--644, 2020.

\bibitem{Siwen2023}
S.~Chen, Y.~Chen, Y.~Huang, and W.~Wei.
\newblock Optimization of {LEACH} routing protocol algorithm.
\newblock In {\em Proc. of the {IEEE} International Conference on Power,
  Electronics and Computer Applications}, pages 1105--1108, 2023.

\bibitem{Mung}
M.~Chiang and T.~Zhang.
\newblock Fog and {IoT}: An overview of research opportunities.
\newblock {\em IEEE Internet of Things Journal}, 3(6):854--864, 2016.

\bibitem{dohare2019pso}
I.~Dohare and K.~Singh.
\newblock {PSO-DEC}: {PSO} based deterministic energy efficient clustering
  protocol for {IoT}.
\newblock {\em Journal of Discrete Mathematical Sciences and Cryptography},
  22(8):1463--1475, 2019.

\bibitem{gamal2022enhancing}
M.~Gamal, N.~Mekky, H.~Soliman, and N.~Hikal.
\newblock Enhancing the lifetime of wireless sensor networks using fuzzy logic
  {LEACH} technique-based particle swarm optimization.
\newblock {\em IEEE Access}, 10:36935--36948, 2022.

\bibitem{gawade2016centralized}
R.~Gawade and S.~Nalbalwar.
\newblock A centralized energy efficient distance based routing protocol for
  wireless sensor networks.
\newblock {\em Journal of Sensors}, 2016, 2016.

\bibitem{gouda2020systematic}
O.~Gouda, A.~Nassif, M.~AbuTalib, and Q.~Nasir.
\newblock A systematic literature review on metaheuristic optimization
  techniques in {WSNs}.
\newblock {\em International journal of mathematics and computers in
  simulation}, 14:187--192, 2020.

\bibitem{ben2022energy}
B.~Gouissem, R.~Gantassi, and S.~Hasnaoui.
\newblock Energy efficient grid based k-means clustering algorithm for large
  scale wireless sensor networks.
\newblock {\em International Journal of Communication Systems}, 35(14):e5255,
  2022.

\bibitem{guhan2021eedchs}
T.~Guhan, N.~Revathy, K.~Anuradha, and B.~Sathyabama.
\newblock {EEDCHS-PSO}: Energy-efficient dynamic cluster head selection with
  differential evolution and particle swarm optimization for wireless sensor
  networks ({WSNS}).
\newblock In {\em Evolution in Computational Intelligence}, pages 715--726.
  2021.

\bibitem{han2022novel}
B.~Han, F.~Ran, J.~Li, L.~Yan, H.~Shen, and A.~Li.
\newblock A novel adaptive cluster based routing protocol for energy-harvesting
  wireless sensor networks.
\newblock {\em Sensors}, 22(4):1564, 2022.

\bibitem{Humaira2022}
H.~Harun, M.~Islam, and M.~Hanif.
\newblock Genetic algorithm for efficient cluster head selection in leach
  protocol of wireless sensor network.
\newblock In {\em Proc. of the International Conference on Advancement in
  Electrical and Electronic Engineering}, pages 1--6, 2022.

\bibitem{Heinzelman2}
W.~Heinzelman, A.~Chandrakasan, and H.~Balakrishnan.
\newblock Energy-efficient communication protocol for wireless microsensor
  networks.
\newblock In {\em Proc. of the 33rd Annual {IEEE} Hawaii International
  Conference on System Sciences ({HICSS}), {Hawaii}, {USA}}, volume~2, pages
  1--10, January 2000.

\bibitem{Heinzelman}
W.~Heinzelman, A.~Chandrakasan, and H.~Balakrishnan.
\newblock An application-specific protocol architecture for wireless
  microsensor networks.
\newblock {\em IEEE Transactions on Wireless Communications}, 1(4):660--670,
  2002.

\bibitem{Jamshed22}
M.~Jamshed, K.~Ali, Q.~Abbasi, M.~Imran, and M.~Ur-Rehman.
\newblock Challenges, applications, and future of wireless sensors in internet
  of things: A review.
\newblock {\em IEEE Sensors Journal}, 22(6):5482--5494, 2022.

\bibitem{Jukuntla22}
A.~Jukuntla and V.~Dondeti.
\newblock A comparative analysis on variants of {LEACH} based routing protocols
  for prolonging network lifetime: Survey.
\newblock In {\em Proc. of the International Conference on Computing,
  Communication, and Intelligent Systems ({ICCCIS})}, pages 351--355, 2022.

\bibitem{Juwaid2018}
A.~Juwaied and L.~Jackowska-Strumitto.
\newblock Analysis of cluster heads positions in stable election protocol for
  wireless sensor network.
\newblock In {\em Proc. of the International Interdisciplinary {PhD} Workshop
  ({IIPhDW})}, pages 367--370, 2018.

\bibitem{kandari2014k}
B.~Kandari and R.~Singh.
\newblock {K-SEP}: {A} more stable {SEP} using {K-Means} clustering and
  probabilistic transmission in {WSN}.
\newblock {\em International Journal of Current Engineering and Technology},
  4(4), 2014.

\bibitem{kaur21}
G.~Kaur, P.~Chanak, and M.~Bhattacharya.
\newblock Energy-efficient intelligent routing scheme for {IoT}-enabled {WSNs}.
\newblock {\em IEEE Internet of Things Journal}, 8(14):11440--11449, 2021.

\bibitem{Kumawat}
S.~Kumawat, H.~Kaur, and O.~Dahiya.
\newblock A modified {LEACH} protocol in wireless sensor networks for energy
  efficient routing protocol.
\newblock In {\em Proc. of the International Conference on Intelligent
  Engineering and Management ({ICIEM})}, pages 785--790, 2022.

\bibitem{lata2020fuzzy}
S.~Lata, S.~Mehfuz, S.~Urooj, and F.~Alrowais.
\newblock Fuzzy clustering algorithm for enhancing reliability and network
  lifetime of wireless sensor networks.
\newblock {\em IEEE Access}, 8(4):66013--66024, 2020.

\bibitem{lehsaini2018improved}
M.~Lehsaini and M.~Benmahdi.
\newblock An improved k-means cluster-based routing scheme for wireless sensor
  networks.
\newblock In {\em Proc. of the International Symposium on Programming and
  Systems ({ISPS})}, pages 1--6, 2018.

\bibitem{lim2011adaptive}
J.~Lim and C.~Bleakley.
\newblock Adaptive {WSN} scheduling for lifetime extension in environmental
  monitoring applications.
\newblock {\em International Journal of Distributed Sensor Networks},
  8(1):286981, 2011.

\bibitem{Jie}
J.~Lin, W.~Yu, N.~Zhang, X.~Yang, H.~Zhang, and W.~Zhao.
\newblock A survey on internet of things: Architecture, enabling technologies,
  security and privacy, and applications.
\newblock {\em IEEE Internet of Things Journal}, 4(5):1125--1142, 2017.

\bibitem{lin2018efficient}
L.~Lin, L.~Donghui, and L.~Ding.
\newblock An efficient routing algorithm based on {K-means++} clustering and
  fuzzy for wireless sensor network.
\newblock In {\em Proc. of the 13th World Congress on Intelligent Control and
  Automation ({WCICA})}, pages 716--720, 2018.

\bibitem{mahboub2017energy}
A.~Mahboub and M.~Arioua.
\newblock Energy-efficient hybrid k-means algorithm for clustered wireless
  sensor networks.
\newblock {\em International Journal of Electrical and Computer Engineering},
  7(4):2054, 2017.

\bibitem{Junqi2022}
J.~Mao, M.~Gu, and Y.~Huo.
\newblock Improved routing algorithm for wireless sensor networks based on
  {LEACH}.
\newblock In {\em Proc. of the International Communication Engineering and
  Cloud Computing Conference}, pages 33--36, 2022.

\bibitem{mechta2014leach}
D.~Mechta, S.~Harous, I.~Alem, and D.~Khebbab.
\newblock {LEACH-CKM}: Low energy adaptive clustering hierarchy protocol with
  {K-means} and {MTE}.
\newblock In {\em Proc. of the 10th International Conference on Innovations in
  Information Technology ({IIT})}, pages 99--103, 2014.

\bibitem{mishra2019trust}
M.~Mishra, G.~S. Gupta, and X.~Gui.
\newblock Trust-based cluster head selection using the {K-Means} algorithm for
  wireless sensor networks.
\newblock In {\em Proc. of the International Conference on Smart Systems and
  Inventive Technology ({ICSSIT})}, pages 819--825, 2019.

\bibitem{Ogundile}
O.~Ogundile and A.~Alfa.
\newblock A survey on an energy-efficient and energy-balanced routing protocol
  for wireless sensor networks.
\newblock {\em Sensors}, 17(5), 2017.

\bibitem{pal2020eewc}
R.~Pal, S.~Yadav, R.~Karnwal, and A.~Aarti.
\newblock {EEWC}: energy-efficient weighted clustering method based on genetic
  algorithm for {HWSNs}.
\newblock {\em Complex \& Intelligent Systems}, 6(2):391--400, 2020.

\bibitem{panchal2020rch}
A.~Panchal, L.~Singh, and R.~Singh.
\newblock {RCH-LEACH}: Residual energy based cluster head selection in {LEACH}
  for wireless sensor networks.
\newblock In {\em Proc. of the {IEEE} International Conference on Electrical
  and Electronics Engineering ({ICE3})}, pages 322--325, 2020.

\bibitem{periyasamy2016balanced}
S.~Periyasamy, S.~Khara, and S.~Thangavelu.
\newblock Balanced cluster head selection based on modified k-means in a
  distributed wireless sensor network.
\newblock {\em International Journal of Distributed Sensor Networks}, 2016:11,
  2016.

\bibitem{pitchaimanickam2020hybrid}
B.~Pitchaimanickam and G.~Murugaboopathi.
\newblock A hybrid firefly algorithm with particle swarm optimization for
  energy efficient optimal cluster head selection in wireless sensor networks.
\newblock {\em Neural Computing and Applications}, 32(12):7709--7723, 2020.

\bibitem{Pundir20}
S.~Pundir, M.~Wazid, D.~Singh, A.~Das, J.~Rodrigues, and Y.~Park.
\newblock Intrusion detection protocols in wireless sensor networks integrated
  to internet of things deployment: Survey and future challenges.
\newblock {\em IEEE Access}, 8:3343--3363, 2020.

\bibitem{Pushpa}
G.~Pushpa, N.~Dharani, and M.~Bennet.
\newblock Optimization of energy in wireless sensor network using leach
  protocol.
\newblock In {\em Proc. of the International Conference on Communication and
  Electronics Systems ({ICCES})}, pages 576--581, 2022.

\bibitem{rao2017particle}
P.~Rao, P.~Jana, and H.~Banka.
\newblock A particle swarm optimization based energy efficient cluster head
  selection algorithm for wireless sensor networks.
\newblock {\em Wireless networks}, 23(7):2005--2020, 2017.

\bibitem{razaque2016h}
A.~Razaque, S.~Mudigulam, K.~Gavini, F.~Amsaad, M.~Abdulgader, and G.~Krishna.
\newblock {H-LEACH}: Hybrid-low energy adaptive clustering hierarchy for
  wireless sensor networks.
\newblock In {\em Proc. of the {IEEE} Long Island Systems, Applications and
  Technology Conference ({LISAT})}, pages 1--4, 2016.

\bibitem{Reddy2023}
M.~Reddy, M.~Chandra, P.~Venkatramana, and R.~Dilli.
\newblock Energy-efficient cluster head selection in wireless sensor networks
  using an improved grey wolf optimization algorithm.
\newblock {\em Computers}, 12(2), 2023.

\bibitem{SinghQoS2022}
O.~Singh, M.~Yadav, P.~Yadav, V.~Rishiwal, D.~Jat, and P.~Thakur.
\newblock {QoS}-attentive learning-based routing for scalable {WSNs}.
\newblock In {\em Proc. of the International Conference on Data Science,
  Machine Learning and Artificial Intelligence}, page 305–310, 2022.

\bibitem{smail2017energy}
O.~Smail, B.~Cousin, and I.~Snoussaoui.
\newblock Energy-aware and stable cluster-based multipath routing protocol for
  wireless ad hoc networks.
\newblock {\em International Journal of Networking and Virtual Organisations},
  17(2), 2017.

\bibitem{SEP_SANPA04}
G.~Smaragdakis, I.~Matta, and A.~Bestavros.
\newblock {SEP}: A stable election protocol for clustered heterogeneous
  wireless sensor networks.
\newblock In {\em Proc. of the 2nd International Workshop on Sensor and Actor
  Network Protocols and Applications ({SANPA}), {Massachusetts}, {USA}}, pages
  1--11, August 2004.

\bibitem{suleiman2022cost}
H.~Suleiman.
\newblock A cost-aware framework for {QoS}-based and energy-efficient
  scheduling in cloud--fog computing.
\newblock {\em Future Internet}, 14(11):333, 2022.

\bibitem{suleiman2021adaptP}
H.~Suleiman and M.~Hamdan.
\newblock Adaptive probabilistic model for energy-efficient distance-based
  clustering in {WSNs} ({Adapt-P}): A {LEACH}-based analytical study.
\newblock {\em {Journal of Wireless Mobile Networks, Ubiquitous Computing, and
  Dependable Applications (JoWUA)}}, 12(3):65--86, 2021.

\bibitem{wu2022dual}
M.~Wu, Z.~Li, J.~Chen, Q.~Min, and T.~Lu.
\newblock A dual cluster-head energy-efficient routing algorithm based on
  canopy optimization and {K-Means} for {WSN}.
\newblock {\em Sensors}, 22(24):9731, 2022.

\bibitem{Daogen2022}
X.~Yin, L.~Guan, and D.~Jiang.
\newblock A survey on {LEACH}-based clustering routing protocols in wireless
  sensor networks.
\newblock In {\em Proc. of the International Conference on Education, Network
  and Information Technology}, pages 323--325, 2022.

\bibitem{Zafor22}
H.~Zafor, N.~Mazumdar, and A.~Nag.
\newblock A comparative study of survey papers based on energy efficient,
  coverage-aware, and fault tolerant in static sink node of {WSN}.
\newblock In {\em Proc. of the {IEEE} International Conference on Electrical,
  Electronics and Computer Engineering ({UPCON})}, pages 1--6, 2022.

\bibitem{Zagrouba21}
R.~Zagrouba and A.~Kardi.
\newblock Comparative study of energy efficient routing techniques in wireless
  sensor networks.
\newblock {\em Information}, 12(1), 2021.

\bibitem{zhang2017fuzzy}
Y.~Zhang, J.~Wang, D.~Han, H.~Wu, and R.~Zhou.
\newblock Fuzzy-logic based distributed energy-efficient clustering algorithm
  for wireless sensor networks.
\newblock {\em Sensors}, 17(7):1554, 2017.

\end{thebibliography}
